\documentclass[12pt]{article}
\usepackage{fullpage, amssymb, amsmath, amsthm}
\usepackage{color}

\numberwithin{equation}{section}

\newtheorem{theorem}[equation]{Theorem}

\newtheorem{lemma}[equation]{Lemma}

\newtheorem*{restatedmainthm}{Theorem 1.1 (restated)}

\DeclareMathOperator{\GDM}{GDM}

\newcommand{\bool}[1]{\{0, 1\}^{#1} }
\newcommand{\suc}[2]{\success_{#1, #2}}

\DeclareMathOperator{\zerr}{0-err}

\DeclareMathOperator{\close}{\textbf{close}}

\DeclareMathOperator{\seeds}{\textbf{seeds}}
\DeclareMathOperator{\states}{\textbf{states}}
\DeclareMathOperator{\queries}{\textbf{queries}}

\DeclareMathOperator{\success}{Suc}

\DeclareMathOperator{\supp}{supp}
\DeclareMathOperator{\val}{Val}

\DeclareMathOperator{\size}{size}

\newcommand{\eps}{\varepsilon}

\newcommand{\bE}{\mathbb{E}}

\newcommand{\bR}{\mathbb{R}}

\begin{document}

%%\ccsps{}

%%% personal macros and packages
%%% end of personal macros and packages

\title{Improved Direct Product Theorems for Randomized Query Complexity}% Insert title here. (Use \\ to split lines.)
\author{Andrew Drucker\thanks{Institute for Advanced Study, Princeton, NJ, USA. Email:  \ andy.drucker@gmail.com.  This research was conducted while the author was a graduate student at MIT, with support from a DARPA YFA grant of Scott Aaronson.  This version essentially matches the journal version (Computational Complexity 21(2), pp. 197-244, 2012) for which copyright is held by Springer.  The final publication is available at springerlink.com.
 }}% Insert author list here. (Each author must be given
         % including his/her current address, email and possibly
         % homepage.  Lines are seperated by \\, different authors
         % are seperated by \and .)
         
\date{}      
         
       \maketitle

%%\authorhead{}% If necessary insert running head authors here.
%%\authorlist{}% If necessary insert title author list here.

\begin{abstract}
The ``direct product problem'' is a fundamental question in complexity theory which seeks to understand how the difficulty of computing a function on each of $k$ independent inputs scales with $k$.
We prove the following direct product theorem (DPT) for query complexity: if every $T$-query algorithm
has success probability at most $1 - \eps$ in computing the Boolean function $f$ on input distribution $\mu$, then for $\alpha \leq 1$, every $\alpha \eps Tk$-query algorithm has success probability at most $(2^{\alpha \eps}(1-\eps))^k$ in computing the $k$-fold direct product $f^{\otimes k}$ correctly on $k$ independent inputs from $\mu$.  In light of examples due to Shaltiel, this statement gives an essentially optimal tradeoff between the query bound and the error probability.  Using this DPT, we show that for an absolute constant $\alpha > 0$, the worst-case success probability of any $\alpha R_2(f) k$-query randomized algorithm for $f^{\otimes k}$ falls exponentially with $k$.  The best previous statement of this type, due to Klauck, \v{S}palek, and de Wolf, required a query bound of $O(bs(f)  k)$.

Our proof technique involves defining and analyzing a collection of martingales associated with an algorithm attempting to solve $f^{\otimes k}$.  Our method is quite general and yields a new XOR lemma and threshold DPT for the query model, as well as DPTs for the query complexity of learning tasks, search problems, and tasks involving interaction with dynamic entities.  We also give a version of our DPT in which decision tree size is the resource of interest.\end{abstract}

%\begin{keywords}
%direct product theorems; query complexity; decision trees; average-case complexity; hardness amplification
%\end{keywords}

%\begin{subject}
%68Q17  %using   http://www.ams.org/mathscinet/msc/msc2010.html?t=68-XX&btn=Current
%\end{subject}

\section{Introduction}

\subsection{Direct product theorems}
Suppose some Boolean function $f(x)$ on $n$ input bits is ``hard to compute'' for a certain computational model.  It seems that computing the $k$-tuple $f^{\otimes k}(x^1, \ldots, x^k) := (f(x^1), \ldots, f(x^k))$ on independent inputs $x^1, \ldots, x^k$ should be ``even harder.''  The intuition is that the $k$ tasks to be performed appear separate and unrelated, and that with more tasks one is more likely to make a mistake.  
One way to make this idea more precise is to study the \emph{direct product problem}, in which we try to prove statements of the following form:

\textit{Suppose every algorithm using resources at most $T$ has success probability at most $ p$ in computing $f$.  Then, every algorithm using resources at most $T'$ has success probability at most $p'$ in computing $f^{\otimes k}$ on $k$ independent inputs to $f$.}

Such a result is called a \emph{direct product theorem (DPT)}.  The direct product problem can be contrasted with a second, related question, the \emph{direct sum problem}, which studies how the complexity of solving $k$ instances of a problem scales with $k$, when we are only interested in algorithms which succeed with high probability (or probability 1).  For a recent overview of the direct sum problem in query complexity, and proofs of some new results, see~\cite{JKS}.

Depending on the computational model and our interests, $T$ and $T'$ might measure time, communication, or any other resource.  The success probability could be with respect to some input distribution $\mu$, in which case it is natural to assume in the $k$-fold setting that the inputs are drawn independently from $\mu$; we call this the \emph{average-case} setting.  However, one can also consider the case where $p$ is a bound on the \emph{worst-case} success probability of a randomized algorithm, ranging over all inputs to $f$; we then try to establish an upper-bound $p'$ on the worst-case success probability of query-bounded algorithms for $f^{\otimes k}$.  
The strength of a direct product theorem can be measured in terms of the dependence of the parameters $T', p'$ on $T, p, k$, and, possibly, on the function $f$ itself.  We want $T'$ to be large and $p'$ to be small, to establish that the $k$-fold problem is indeed ``very hard.''

There is also an important variant of the direct product problem, in which we are interested in computing the ``$k$-fold XOR'' $f^{\oplus k}(x^1, \ldots, x^k) := f(x^1) \oplus \ldots \oplus f(x^k)$ of $k$ independent inputs to $f$.  An \emph{XOR lemma} is a result which upper-bounds the success probability $p'$ achievable by algorithms for $f^{\oplus k}$ using $T'$ resources, under the assumption that any algorithm using $T$ resources has success probability at most $p$.\footnote{Terminology varies somewhat in the literature.  For instance, what we call XOR lemmas are called ``direct product theorems'' in \cite{Sha}, and what we refer to as direct product problems are called the ``concatenation variant'' by Shaltiel.}  An obvious difference from DPTs is that in an XOR lemma, $p'$ must always be at least $1/2$, since $f^{\oplus k}$ is Boolean and the algorithm could simply guess a random bit.  The hope is that $(p' - 1/2)$ decays exponentially with $k$.  Research on XOR lemmas has proceeded in parallel with research on direct product theorems; the known results are of similar strength (with some exceptions), and in some cases there are reductions known from XOR lemmas to DPTs or vice versa; see~\cite{Ung, IK10} for an overview and recent results of this type.

The direct product problem has been studied extensively in models such as Boolean circuits, e.g., in~\cite{GNW, IW,IJKW10}; communication protocols~\cite{IRW, Sha, KSdW, LSS, VW}; and query algorithms~\cite{IRW, NRS,Sha,KSdW}.  
In all of these models, an optimal $T$-bounded algorithm which attempts to compute $f$ can always be applied independently to each of $k$ inputs, using at most $T' = Tk$ resources and succeeding with probability $p' = p^k$, so these are the ``ideal,'' strongest parameters one might hope for in a DPT.   However, direct product statements of such strength are generally false, as was shown by~\cite{Sha}, who gave a family of counterexamples which applies to all ``reasonable'' computational models.  We will describe these examples (specialized to the query model) in Section~\ref{examplesec}.\footnote{Shaltiel calls a DPT ``strong'' if it applies to all $p, T$ and its parameters satisfy $p' \leq  p^{\Omega(k)}$ and $T' \geq \Omega(Tk)$.  His counterexamples rule out strong DPTs for most computational models.  In later works, the modifier ``strong'' has been used in a somewhat broader way.  We will not use this terminology in the present paper.}

Thus, all DPTs shown have necessarily been weaker in one of several ways.  First, researchers have restricted attention to algorithms of a special form. \cite{Sha} showed a DPT with the ``ideal'' parameters above holds for the query model, if the algorithm is required to query each of the $k$ inputs exactly $T$ times.  He called such algorithms ``fair.''\footnote{Actually, Shaltiel proved, in our terms, an optimal XOR lemma for fair algorithms, but as he noted, this implies an optimal DPT, and his proof method can also be modified to directly prove an optimal DPT for fair algorithms.}  
A similar result for a special class of query algorithms called ``decision forests'' was shown earlier in \cite{NRS}.

Second, DPTs have been shown for unrestricted algorithms, but using resource bounds whose strength depends on properties of the function $f$. These results require the resource bound $T'$ to scale as $\mathcal{D}(f) k$, where $\mathcal{D}(f)$ is some complexity measure that can be significantly smaller than the resources needed to compute a single instance of $f$.
For example, \cite{KSdW} showed that for any $f$ and any $\gamma > 0$, a DPT holds for $f$ in which the achievable worst-case success probability $p'$ is at most $\left(1/2 + \gamma \right)^k$, provided $T' \leq \alpha \cdot bs(f) k$ for some constant $\alpha = \alpha(\gamma) > 0$.  Here $bs(f)$ is the \emph{block sensitivity} of $f$ \cite{Nis89, BdW}, a complexity measure known to be related to the randomized query complexity by the inequalities $R_2(f)^{1/3} \leq bs(f) \leq R_2(f)$ (suppressing constant factors). Now, one can always compute $f$ correctly on $k$ instances with high probability using $O(R_2(f) k \log k)$ queries.  For many functions, including random functions, $bs(f) = \Theta(R_2(f))$ so in these cases the DPT of~\cite{KSdW} gives a fairly tight result.  However, examples are known~\cite{BdW} where $bs(f) = O(\sqrt{R_2(f)})$, so the number of queries allowed by this DPT can be significantly less than one might hope.

Klauck, \v{S}palek, and de Wolf also proved DPTs for \emph{quantum} query algorithms computing $f$, in which the worst-case success probability $p'$ drops exponentially in $k$ if the number of allowed quantum queries is $O(\sqrt{bs(f)} k)$.  For symmetric functions, direct product theorems of a strong form were proved for quantum query complexity by \cite{ASdW}.  \cite{Spa} proved a DPT for quantum query algorithms where the resource bound $T'$ scales in terms of a complexity measure called the \emph{multiplicative quantum adversary}.  
After a preprint of our paper appeared, a sequence of works~\cite{Sher, AMRR, LR11} dramatically advanced our understanding of the direct product problem in the quantum query model.  This culminated in a DPT for quantum queries \cite{LR11} in which the success probability decays exponentially even as the query bound scales as $\Omega(Q_2(f) k)$.  Here, $Q_2(f)$ is the bounded-error quantum query complexity of a (possibly non-Boolean) function $f$.

In the model of communication protocols, several types of results have been shown.  DPTs have been given for specific functions: e.g., in~\cite{KSdW} a DPT was proved for the quantum communication complexity of the Disjointness function, and a classical analogue was proved by \cite{Kla10}.  On the other hand, general DPTs have been given, whose resource bound scales in terms of complexity measures that may be significantly smaller than the communication complexity of $f$.
For example, in communication complexity, DPTs have been shown whose strength is related to the so-called \emph{discrepancy} of $f$~\cite{Sha, LSS}.
 
%%%%

Since the present work, there has been significant progress in the communication model.  In the public-coin randomized setting, \cite{Jain11} showed a strong general-purpose DPT for one-way communication, and new DPTs were shown for two-way communication in~\cite{Jain11, JPY12}.  Sherstov~\cite{Sher} gave a new DPT for quantum communication, whose resource bound scales as $\Omega(\GDM(f)k)$, where $\GDM(f)$ is the lower bound on quantum communication complexity obtained by the \emph{generalized discrepancy method}---the strongest lower-bound technique known in the quantum setting.

In the Boolean circuit model, despite intensive study, the known results are quantitatively much weaker, and in particular require $T'$ to \emph{shrink} as $k$ grows in order to make the success probability $p'$ decay as $k$ grows.  It is at least known that, under this limitation, a DPT with $p' = p^k$ can be shown using~\cite{Imp, Hol05}, as remarked in \cite{IK10}.

\subsection{Our results}

Our first result is the following direct product theorem in the average-case setting:

\begin{theorem}\label{mainsdpt}  Suppose $f$ is a Boolean function and $\mu$ is a distribution over inputs to $f$, such that any $T$-query randomized algorithm has success probability at most $(1 - \eps)$ in computing $f$ on an input from $\mu$.  Then for $0 < \alpha \leq 1$, any randomized algorithm making $\alpha \eps T k$ queries  has success probability at most $\left(2^{\alpha \eps}(1 - \eps)\right)^k < (1 - \eps + .84\alpha\eps)^k$ in computing $f^{\otimes k}$ correctly on $k$ inputs drawn independently from $\mu$.
\end{theorem}
We use Shaltiel's examples to show that the tradeoff in Theorem~\ref{mainsdpt} between the query bound and the error probability is essentially best-possible, at least for general functions $f$ and for small values $\alpha < .01$.  (For \emph{specific} functions, the success probability will in some cases decay exponentially even when the number of queries allowed scales as $T k$ rather than $\eps T k$.) Theorem~\ref{mainsdpt} reveals that small values of $\eps$, as used in Shaltiel's examples, are the only major ``obstruction'' to strong, general direct product statements in the query model.

Using Theorem~\ref{mainsdpt}, we obtain the following DPT for worst-case error, which strengthens the worst-case DPT of \cite{KSdW} mentioned earlier:

\begin{theorem}\label{worstcasesdpt} For any Boolean function $f$ and $0 < \gamma< 1/4$, any randomized algorithm making at most $\gamma^3 R_2(f) k/ 11$ queries has worst-case success probability less than $\left( 1/2 + \gamma \right)^k$ in computing $f^{\otimes k}$ correctly.
\end{theorem}

It seems intuitive that some statement like Theorem~\ref{worstcasesdpt} should hold, and proving such a DPT was arguably one of the major open problems in classical query complexity.\footnote{While classical query algorithms can be viewed as a subclass of quantum query algorithms, we note that Theorem~\ref{worstcasesdpt} is incomparable to the more-recent quantum DPT proved in \cite{LR11}: our result shows exponentially-decaying success probability for a more restricted class of algorithms, but under a potentially larger query bound.}

We also prove a new XOR lemma.  Let $B_{k, p}$ denote the binomial distribution on $k$ trials with success probability $p$.

\begin{theorem}\label{xorsdpt}  Suppose that any $T$-query randomized algorithm has success probability at most $(1 - \eps)$ in computing the Boolean function $f$ on an input from $\mu$.  Then for $0 < \alpha \leq 1$, any randomized algorithm making $\alpha \eps T k$ queries and attempting to compute $f^{\oplus k}$ on $k$ inputs drawn independently from $\mu$ has success probability at most 
\[\frac{1}{2} \left(1 + \Pr_{Y \sim B_{k, 1 - 2\eps}}[Y > (1 - \alpha\eps)k] \right), \]
which is less than $\frac{1}{2} \left(1 + \left[ 1 - 2\eps + 6\alpha \ln(2/\alpha) \eps  \right]^{k}\right)$.
\end{theorem}

Compare the probability bound above with the success probability $\frac{1}{2}(1 + (1 - 2\eps)^k)$, which can be attained using $Tk$ queries by attempting to solve each instance independently and outputting the parity of the guessed bits.  The concrete estimate given in Theorem~\ref{xorsdpt} is meant to illustrate how our bound approaches this value as $\alpha \rightarrow 0$.  By a more careful use of Chernoff inequalities, one can get somewhat tighter bounds for specific ranges of $\alpha, \eps$.  An XOR lemma for the worst-case setting can also be derived from our result.

In addition to our ``ordinary'' DPT (Theorem~\ref{mainsdpt}), we also prove a ``threshold'' DPT, which bounds the probability that a query-bounded algorithm for $f^{\otimes k}$ solves ``many'' of the $k$ instances correctly.  As one special case, we prove:

\begin{theorem}\label{special_tsdpt}  Let $f$ be a (not necessarily Boolean) function such that any $T$-query algorithm has success probability at most $1 - \eps$ in computing $f$ on an input from $\mu$.  Fix $\eta, \alpha \in (0, 1]$.  Consider any randomized algorithm $\mathcal{R}$ making at most $\alpha \eps Tk$ queries on $k$ independent inputs from $\mu$.  The probability that $\mathcal{R}$ computes $f$ correctly on at least $\eta k$ of the inputs is at most
\[    \Pr_{Y \sim B_{k, 1 - \eps}}[ Y \geq (\eta - \alpha \eps) k].        \]
\end{theorem}

Using Chernoff inequalities, Theorem~\ref{special_tsdpt} gives success bounds which decay exponentially in $k$ for any fixed $\alpha, \eps, \eta$, provided $\eta > 1 - \eps + \alpha \eps$.  As we will explain, Shaltiel's examples show that this cutoff is nearly best-possible.
By setting $\eta := 1$ in Theorem~\ref{special_tsdpt}, we also get an ordinary DPT for non-Boolean functions, which for typical parameter settings is stronger than the DPT we'd obtain by a straightforward generalization of our techniques for Theorem~\ref{mainsdpt}.  This is the simplest way we know to get such a DPT.

Threshold DPTs have been proved for a variety of models, including, recently, for arbitrary Boolean functions in the quantum query model \cite{LR11}.  \cite{Ung} showed how to derive threshold DPTs from XOR lemmas, and recent work of \cite{IK10} gave a way to derive threshold DPTs from sufficiently strong DPTs; see also the earlier works cited in~\cite{Ung, IK10}.  
However, the results of \cite{IK10} do not apply for our purposes, and the threshold DPT we prove is more general than we'd get by applying the results of \cite{Ung} to our XOR lemma.  In any case the proof of our threshold DPT is, we feel, quite natural, and actually forms the basis for the proof of our XOR lemma.  
Our method for proving threshold DPTs applies to very general threshold events: we give bounds on the probability that the set $S \subseteq [k]$ of instances solved correctly by a query-bounded algorithm is ``large,'' in a sense specified by an arbitrary monotone collection $\mathcal{A}$ of subsets of $[k]$.  Generalized threshold DPTs of this form were shown recently by \cite{HS11} in the circuit model, for a rich class of computational tasks called ``weakly verifiable puzzles;'' as usual in the circuit model, these DPTs require $T'$ to shrink with $k$.  Our techniques appear unrelated to theirs.

We also prove new DPTs for relations (for which direct \emph{sum} theorems were proved recently by \cite{JKS}), learning tasks, search problems, and errorless heuristics.  Deterministic query algorithms can be equivalently viewed as \emph{decision trees}, and we also prove a DPT for decision trees in which decision tree \emph{size}, rather than depth (i.e., number of queries), is the resource of interest.  Impagliazzo, Raz, and Wigderson~\cite{IRW} gave a DPT for decision tree size with ``ideal'' success probability decay $p' = p^k$, but in the case where the size is not allowed to scale with $k$, i.e., the setting $T' = T$.  By contrast, in our DPT,  the success probability decays as $p^{\Omega(k)} = (1 - \eps)^{\Omega(k)}$, while the size bound $T'$ scales as $T^{\Omega(\eps k)}$.  

Finally, we give a further generalization of our DPTs, in which the $k$ objects being queried are \emph{dynamic entities} rather than static strings---that is, the answers to current queries may depend on past queries.  DPTs for dynamic interaction have been proved before \cite{MPR07}, but only for the case in which the number of queries to each entity is fixed in advance.  (This is analogous to Shaltiel's result for ``fair'' algorithms.)  We further discuss the relation to past work on dynamic interaction in Section~\ref{generalsec}.

In order to ease notation, in this paper we discuss only DPTs for total functions, but our results apply to partial functions, that is, functions with a restricted domain; the proofs are the same.  Similarly, our theorems and proofs carry over without change to handle non-Boolean input alphabets, as well as heterogeneous query costs.
Taken as a whole, our results provide a fairly complete picture of the ``direct product phenomenon'' for randomized query complexity, although there may still be room for improvement in some of our bounds.
We hope this work may also help lead to a better understanding of the direct product problem in other, richer computational models.

\subsection{Our methods}

We first explain our method to prove our ``basic'' direct product theorem, Theorem~\ref{mainsdpt}.  As mentioned earlier, Shaltiel~\cite{Sha} proved an optimal DPT for ``fair'' decision trees, in which each of the $k$ inputs receives $T$ queries.  Our proof method for Theorem~\ref{mainsdpt} also yields an alternate proof of Shaltiel's result, and it is helpful to sketch how this works first.  (Really, this ``alternate proof'' is little more than a rephrasing of Shaltiel's proof technique, but the rephrasing gives a useful perspective which helps us to prove our new results.)

Suppose that every $T$-query algorithm for computing $f$ succeeds with probability at most $1 - \eps$ on an input from the distribution $\mu$.  Consider a fair $Tk$-query algorithm $\mathcal{D}$ for $f^{\otimes k}$, running on $k$ independent inputs from $\mu$.  We think of the algorithm as a ``gambler'' who bets at $k$ ``tables,'' and we define a random variable $X_{j, t} \in [1/2, 1]$ which represents the gambler's ``fortune'' at the $j$-th table after $\mathcal{D}$ has made $t$ queries overall to the $k$ inputs.   Roughly speaking, $X_{j, t}$ measures how well the algorithm is doing in determining the value of $f$ on the $j$-th input.  When $\mathcal{D}$ queries the $j$-th input, the $j$-th fortune may rise or fall, according to the bit seen; we regard each bit revealed to be generated sequentially at random, conditioned on the bits queried so far.  The fortunes are defined so that $X_{j, 0} \leq 1 - \eps$ for each $j$ (reflecting the assumed hardness of $f$ on $\mu$), and so that no action by the algorithm leads to an expected gain in fortune.\footnote{In standard probabilistic terms, each individual sequence $X_{j, 0}, X_{j, 1}, \ldots$ is a \emph{supermartingale}.  We will not use this terminology in the paper.}  It follows that $\bE[\prod_{j \in [k]}X_{j, Tk}] \leq (1 - \eps)^k$.  But the fortunes are defined so that $\bE[\prod_{j \in [k]}X_{j, Tk}]$ upper-bounds the success probability of $\mathcal{D}$ in computing $f^{\otimes k}$. This gives the DPT for fair algorithms.  A key fact underlying the success of this proof strategy is that, after conditioning on any initial sequence of outcomes to the first $t \leq T$ queries by the algorithm, the $k$ inputs remain independent.  

If $\mathcal{D}$ is no longer required to be fair, but instead makes at most $\alpha \eps Tk$ queries, then the individual fortune $X_{j, t}$ we define no longer has the same intuitive meaning after the $j$-th input has been queried more than $T$ times.  (In this event we simply set $X_{j, t}$ to $1/2$, so that the gambler cannot hope to increase the $j$-th fortune.)  However, the success probability of $\mathcal{D}$ can still be upper-bounded by $\bE[\prod_{j \in S}X_{j, \alpha\eps Tk}]$, where $S$ is the (random) set of inputs which receive at most $T$ queries.  Counting tells us that fewer than $\alpha\eps k$ of the inputs can lie outside of $S$, and each fortune is always at least $1/2$, so the success probability is at most $2^{\alpha\eps k} \bE[\prod_{j \in [k]}X_{j, \alpha \eps Tk}] \leq 2^{\alpha \eps k}(1 - \eps)^k$, giving the statement of Theorem~\ref{mainsdpt}.

Our worst-case DPT for Boolean functions follows straightforwardly from Theorem~\ref{mainsdpt}, by an application of Yao's minimax principle.  Our DPT for decision tree size requires a somewhat different analysis, in which we track the ``size-usage'' of each of the $k$ inputs rather than their number of queries, but the basic approach is the same as in Theorem~\ref{mainsdpt}.  In generalizing our method to prove our other results, however, we face a new wrinkle: the natural definitions of the ``fortunes'' $X_{j, t}$ in these settings are no longer bounded from below by $1/2$.  For example, if $f: \{0, 1\}^n \rightarrow B$ then we have $X_{j, t} \geq |B|^{-1}$, and a straightforward modification of the method described above gives a DPT whose strength degrades as $|B|$ grows.  In other settings (e.g., the $k$-fold XOR setting), we will only have $X_{j, t} \geq 0$, and the method fails completely.\footnote{One way to work around the problem is to simply add a small ``buffer term'' to the fortunes $X_{j, t}$.  However, this leads to poorer bounds, and does not yield our generalized threshold DPTs.}

To overcome this difficulty, we adopt a more general perspective.  Our previous proof hinged on the fact that, if a gambler plays neutral or unfavorable games at $k$ tables with an initial (nontransferable) endowment of $1 - \eps$ at each table, then the probability he reaches a fortune of 1 at every table is at most $(1 - \eps)^k$.  Note, this is just the success probability he would achieve if he followed an independent ``all-or-nothing bet'' strategy at each table.  It is natural to wonder whether this strategy remains optimal if the gambler wants merely to reach a fortune of 1 at ``sufficiently many'' tables.  Indeed, we prove (by an induction on the number of rounds of gambling) that this is true, where the meaning of ``sufficiently many'' can be specified by any monotone collection of subsets of $[k]$.  Most of our generalizations of Theorem~\ref{mainsdpt}, as well as our XOR lemma, follow readily from this handy ``gambling lemma,'' although care is required to define the correct fortunes in each case.

\subsection{Organization of the paper}

In Theorem~\ref{maindefsec} we review preliminaries that are used throughout the paper and that are needed to state and prove our ``basic'' DPTs, Theorems~\ref{mainsdpt} and \ref{worstcasesdpt}.  We will introduce other definitions as needed in later sections.  
In Section~\ref{mainsdptsec} we prove Theorem~\ref{mainsdpt}, and in Section~\ref{examplesec} we use Shaltiel's examples to analyze the tightness of this result. We prove Theorem~\ref{worstcasesdpt} in Section~\ref{worstcasesec}.  

In Section~\ref{gentsec} we prove our ``gambling lemma'' (Lemma~\ref{gamblers}), and use it to prove a generalized threshold DPT for relations. Theorem~\ref{special_tsdpt} will follow as a special case.  We also explain how our threshold DPT implies a DPT for the query complexity of certain learning tasks.  We prove Theorem~\ref{xorsdpt}, our XOR lemma, in Section~\ref{xorsec} (also using Lemma~\ref{gamblers}).
We define search problems and errorless heuristics in Section~\ref{searchsec}, and give DPTs for these settings.  

We prove our DPT for decision tree size in Section~\ref{sizesec}.  In Section~\ref{generalsec}, we describe generalizations of our DPTs to settings involving interaction with dynamic entities.  We end with some questions for future work.

\section{Preliminaries}\label{maindefsec}

All of our random variables will be defined over finite probability spaces.  We let $\supp(X)$ denote the support of a random variable $X$, i.e., the set of values with nonzero probability.
Let $\mu^{\otimes k}$ denote $k$ independent copies of distribution $\mu$.

\subsection{Randomized decision trees and query complexity}

A \emph{decision tree} $\mathcal{D}$ over $\bool{n}$ is a rooted, full binary tree (i.e., each node has either 0 or 2 children), in which interior vertices $v$ are labeled by indices ind$(v) \in [n]$ and leaf vertices are labeled by values $\ell(v)$ in some finite set $B$ (often $B = \bool{}$).  The \emph{height} of $\mathcal{D}$ is the length of the longest descending path in $\mathcal{D}$.  $\mathcal{D}$ defines a function $f_{\mathcal{D}}: \bool{n} \rightarrow B$ in the following way.  On input $x$ we start at the root and follow a descending path through $\mathcal{D}$; at interior node $v$, we pass to the left subchild of $v$ if $x_{ind(v)} = 0$, otherwise we pass to the right subchild of $v$.  When we reach a leaf vertex $v$, we output the value $\ell(v)$.  Any deterministic algorithm to compute $f$ which queries at most $t$ bits of $x$ on any input can be modeled as a height-$t$ decision tree, and we will freely refer to such a tree as a ``$t$-query deterministic algorithm.''

A \emph{randomized decision tree} is a probability distribution $\mathcal{R}$ over deterministic decision trees.  Upon receiving the input $x$, the algorithm samples $\mathcal{D} \sim \mathcal{R}$, then outputs $\mathcal{D}(x)$. (Every randomized query algorithm can be modeled in this fashion.)  We write $\mathcal{R}(x)$ to denote the random variable giving the output of $\mathcal{R}$ on input $x$.  We say that $\mathcal{R}$ is a \emph{$t$-query randomized decision tree} if every decision tree in the support of $\mathcal{R}$ has height at most $t$.

 For $\eps \in [0, 1]$ and a function $f$ (not necessarily Boolean), we say that $\mathcal{R}$ \emph{$\eps$-computes $f$} if for all inputs $x$, $\Pr[\mathcal{R}(x) = f(x)] \geq 1 - \eps$.  Similarly, if $\mu$ is a distribution over inputs $x \in \bool{n}$, we say that $\mathcal{R}$ \emph{$\eps$-computes $f$ with respect to $\mu$} if $\Pr_{x \sim \mu}[\mathcal{R}(x) = f(x)] \geq 1 - \eps$, where the probability is taken over the random sample $x \sim \mu$ and the randomness used by $\mathcal{R}$.

For a function $f: \bool{n} \rightarrow B$, we define $R_2(f)$, the \emph{two-sided-error randomized query complexity of $f$}, as the minimum $t$ for which there exists a $t$-query randomized decision tree which $1/3$-computes $f$.  We define 
\[\suc{T}{\mu}(f) \ := \ 1 - \eps,\] 
where $\eps \geq 0$ is the minimum value for which some $T$-query-bounded randomized algorithm $\mathcal{R}$ $\eps$-computes $f$ with respect to $\mu$.  By standard arguments, this minimum exists, and is attained by a deterministic height-$T$ decision tree.

For $f: \{0, 1\}^n \rightarrow B$ and $k \geq 1$, define $f^{\otimes k}: \{0, 1\}^{kn} \rightarrow B^k$, the \emph{$k$-fold direct product of $f$}, as $f^{\otimes k}(x^1, \ldots, x^k) := (f(x^1), \ldots, f(x^k))$.  If $f$ is Boolean, define the \emph{$k$-fold XOR of $f$} as $f^{\oplus k} (x^1, \ldots, x^k) := f(x^1) \oplus \ldots \oplus f(x^k)$, where $\oplus$ denotes addition mod 2.

\subsection{Binomial distributions and Chernoff bounds}\label{chernsec}

Let $B_{k, p}$ denote the binomial distribution on $k$ trials with bias $p$.  That is, $B_{k, p}$ is distributed as $Y = \sum_{i = 1}^k Y_i$, where the $Y_i$ are independent and $0/1$-valued with $\Pr[Y_i = 1] = p$.  For $s \in \{0, 1, \ldots, k\}$ we have the explicit formula $\Pr[Y = s] = {k \choose s}p^s (1 - p)^{k - s}$.

The following is a general form of Chernoff's inequality:

\begin{lemma}[\cite{DP}, \S 1.3]\label{genchern}  Suppose $Y \sim B_{k, p}$, with $q := 1 - p$.  Then for $t \in [0, q)$,
\[ \Pr\left[Y > (p + t)k  \right] \  \leq \ \left( \left(  \frac{p}{p+t}\right)^{p + t}  \left(\frac{q}{q-t} \right)^{q - t}   \right)^k  .\]
\end{lemma}

The following form of Chernoff's inequality will be more  convenient for us.

\begin{lemma}\label{smallchern}  
Let $\delta \in (0, 1)$, and let $Y \sim B_{k, 1 - \delta}$. If $\beta \in (0, 1/2]$, then
\[ \Pr[Y > (1 - \beta \delta)k] \  < \ \left[ 1 - \delta + 6 \beta \ln(1/\beta) \delta  \right]^{k}.  \] 
\end{lemma}

\begin{proof}  We apply Lemma~\ref{genchern} with $t:= (1 - \beta) \delta$; we find 
\begin{align} \label{eq:init}
   \Pr\left[Y > (1 - \beta\delta) k \right] \ &= \ \Pr[Y > ((1 - \delta) + (1 -\beta)\delta)k] \nonumber  \\
&\leq \ \left( \left(  \frac{1 - \delta}{1 - \beta\delta}\right)^{1 - \beta\delta}  \left(\frac{\delta}{\delta - (1 -\beta) \delta}  \right)^{\delta - (1-\beta) \delta}   \right)^k  \nonumber  \\ 
&= \ \left( \left(  \frac{1 - \delta}{1 - \beta\delta}\right)^{1 - \beta\delta}  \beta^{-\beta \delta}   \right)^k   \nonumber \\ &\leq \  \left( \left(  1 - \delta + 2 \beta \delta\right)^{1 - \beta\delta}  \beta^{-\beta \delta}   \right)^k  ,   
\end{align}
using $\beta \delta \leq 1/2$.

It is easy to verify that $(1 - \delta + 2 \beta \delta) \geq \beta$, so that
\[   \left(  1 - \delta + 2 \beta \delta\right)^{- \beta\delta}  \cdot \beta^{-\beta \delta} \ \leq \  \beta^{-2 \beta \delta} \ = \ e^{2\beta \ln(1/\beta) \delta}.    \]  
Now $2 \beta \ln (1/\beta) \delta \leq 2 / e < .74$.  By convexity of $e^x$, we have $e^x \leq 1 + ((e^{.74} - 1)/.74)\cdot x \leq 1 + 1.49x$ for all $x \in [0, .74]$.  Thus, $e^{2\beta \ln(1/\beta) \delta} \leq 1 + 3 \beta \ln(1/\beta) \delta$.
Combining these facts with Eq.~\ref{eq:init}, we get
\begin{align*} \Pr\left[Y > (1 - \beta\delta) k \right]   \  &\leq  \  \left[ (1 - \delta + 2 \beta \delta)(1 + 3 \beta \ln(1/\beta)\delta)  \right]^{k}   \\ 
&<  \ \left[ 1 - \delta + 6 \beta \ln (1/\beta)\delta   \right]^{k}.
\end{align*}  \end{proof}

The constant $6$ in Lemma~\ref{smallchern} is not best-possible.  To apply the lemma, it is helpful to understand the behavior of the function $h(x) := x \ln(1/x)$.  This function is increasing on $(0, e^{-1}]$, and as $x \rightarrow 0$, $h(x)$ approaches 0 only slightly more slowly than $x$ itself: for an integer $n > 1$ we have 
\[ h\left(\frac{1}{2n \ln n}\right)  \ = \ \frac{1}{2n \ln n}\cdot \ln (2n \ln n)
\ = \ \frac{1}{n}\cdot \frac{\ln (2n \ln n)}{\ln (n^2)}  \ < \ \frac{1}{n} \ .    \]

\section{Proof of Theorem~\ref{mainsdpt}}\label{mainsdptsec}

In this section we prove our ``basic'' direct product theorem:

\begin{restatedmainthm}   Let $f$ be a Boolean function for which $\suc{T}{\mu}(f) \leq 1 - \eps$.  Then for $0 < \alpha \leq 1$, $\suc{\alpha \eps T k}{\mu^{\otimes k}}(f^{\otimes k}) \leq  (2^{\alpha \eps}(1 - \eps))^k <  (1 - \eps + .84\alpha\eps)^k$.
\end{restatedmainthm}

There is no requirement that $T$ be an integer; this will be useful later in proving Theorem~\ref{worstcasesdpt}.
The success bound $(2^{\alpha \eps}(1 - \eps))^k$ above is actually valid for any $\alpha > 0$, but the bound is trivial whenever $\alpha \geq 2$, so we focus attention on a range where the bound is always meaningful.

\begin{proof} The statement is trivial if $T = 0$ or $\eps = 0$, so assume both are positive.  By convexity, it is sufficient to show the statement for deterministic algorithms.  Also, by a standard limiting argument, it is enough to prove this result under the assumption that $\supp(\mu) = \bool{n}$; this ensures that conditioning on any sequence of query outcomes will be well-defined.

Next we set up some notation and concepts relating to the computation of $f$ on a single input; afterward we will apply our work to the direct-product setting.

For a string $u \in \{0, 1, *\}^n$, let the distribution $\mu^{(u)}$ be defined as a sample from $\mu$, conditioned on the event $[x_i = u_i$, $ \forall i$ such that $u_i \in \{0, 1\}]$.  Let $|u|$ denote the number of $0/1$ entries in $u$.  Let $u[x_i \leftarrow b]$ denote the string $u$ with the $i$-th coordinate set to $b$.  In our proof we consider the bits of an input $\mathbf{y} \sim \mu$ to be generated sequentially at random as they are queried.  Thus if an input is drawn according to $\mu$, and $u$ describes the outcomes of queries made so far (with $*$ in the coordinates that have not been queried), we consider the input to be in the ``state'' $\mu^{(u)}$.  If some index $i \in [n]$ is queried next, then the algorithm sees a 0 with probability $\Pr_{\mathbf{y} \sim \mu^{(u)}}[\mathbf{y}_i = 0]$, in which case the input enters state $\mu^{(u[x_i \leftarrow 0])}$; with the remaining probability the algorithm sees a 1 and the input enters state $\mu^{(u[x_i \leftarrow 1])}$.  Clearly this interpretation is statistically equivalent to regarding the input as being drawn from $\mu$ before the algorithm begins (this is the ``principle of deferred decisions'' of probability theory).

For each $u \in \{0, 1, *\}^n$ with $|u| \leq T$, let
\[  W (u) \ :=  \  \suc{T - |u|}{\mu^{(u)}}(f) .  \]
In words, $W(u)$ measures our ``winning prospects'' of computing $f$ on $\mu$, if we begin with a budget of $T$ queries and our first $|u|$ queries reveal the bits described by $u$, and if we follow an optimal strategy thereafter.
Clearly $W^{ }(u) \in [1/2, 1]$, since an algorithm may simply guess a random bit.  We make two more simple claims about this function.

\begin{lemma} \label{Wlem}
1. $W(*^n) \leq 1 - \eps$.  

\vspace{.5 em}

2. For any $u \in \{0, 1, *\}^n$ with $|u| < T$, and any $i \in [n]$, $\bE_{\mathbf{y} \sim \mu^{(u)}}[ W(u[x_i \leftarrow \mathbf{y}_i])  ] \leq W(u)$.  
\end{lemma}

\begin{proof}  1: This is immediate from our initial assumption $\suc{T}{\mu}(f) \leq 1 - \eps$.

\vspace{.5 em}

2: If the $i$-th coordinate has already been queried (i.e., $u_i \in \{0, 1\}$), then $\mathbf{y}_i = u_i$ with probability 1, so $u[x_i \leftarrow \mathbf{y}_i] = u$ and the statement is trivial.  So assume $u_i = *$.  Let $\mathcal{R}_{0}$, $\mathcal{R}_1$ be algorithms making at most $T - (|u| + 1)$ queries and maximizing the success probabilities on $\mu^{(u[x_i \leftarrow 0])}$, $\mu^{(u[x_i \leftarrow 1])}$ respectively.  Thus, the success probability of $\mathcal{R}_b$ is $W(u[x_i \leftarrow b])$.  Consider an algorithm $\mathcal{R}$ which queries $x_i$, then runs $\mathcal{R}_{b}$ if the bit seen is $b$.  $\mathcal{R}$ makes at most $T - |u|$ queries, and the success probability of $\mathcal{R}$ is $\bE_{\mathbf{y} \sim \mu^{(u)}}[ W(u[x_i \leftarrow \mathbf{y}_i])  ]$.  Thus $W(u)$ is at least this value.
\end{proof}

Now we prove the Theorem.  Let $\mathcal{D}$ be any deterministic algorithm making at most $M := \lfloor \alpha \eps T k \rfloor$ queries, and attempting to compute $f^{\otimes k}$ on input strings $(\mathbf{x}^1, \ldots, \mathbf{x}^k) \sim \mu^{\otimes k}$.  For $j \in [k]$ and $0 \leq t \leq M$, let $u_t^{j} \in \{0, 1, *\}^n$ be the random string giving the outcomes of all queries made to $\mathbf{x}^j$ after $\mathcal{D}$ has made $t$ queries (to the entire input).  We need the following simple but important observation:

\begin{lemma}\label{indlem} Condition on any execution of $\mathcal{D}$ for the first $t \geq 0$ steps, with query outcomes given by $u^1_t, \ldots, u^k_t$.  Then the input is in the state $\mu^{(u^1_t)}\times \ldots \times \mu^{(u^k_t)}$.  That is, the $k$ inputs are independent, with $\mathbf{x}^j$ distributed as $\mu^{(u^j_t)}$.
\end{lemma}

\begin{proof}
Fix any $j \in [k]$ and consider any assignment $(x^{j'})_{j' \in [k] \setminus \{j\}}$ of values $x^{j'} \in \{0, 1\}^n$ to the inputs other than the $j$-th input, where $x^{j'}$ extends $u^{j'}_t$ for each $j' \neq j$.  We show that, after conditioning on the query outcomes $u^{1}_t, \ldots, u^k_t$ \emph{and} on the event $[\mathbf{x}^{j'}= x^{j'}$ $\forall j' \neq j]$, the $j$-th input $\mathbf{x}^j$ is distributed according to $\mu^{(u^j_t)}$.  This will prove the Lemma.

Consider each $y \in \{0, 1\}^n$ which extends $u^j_t$.  Now $u_t^1, \ldots, u^k_t$ are, by assumption, a possible description of the first $t$ queries made by $\mathcal{D}$ under \emph{some} input.  Since $\mathcal{D}$ is deterministic, and $(x^1, \ldots, x^{j - 1}, y, x^{j+1}, \ldots,  x^k)$ are consistent with $(u_t^1, \ldots, u^k_t)$, we conclude that $(u_t^1, \ldots, u^k_t)$ also describe the first $t$ queries made by $\mathcal{D}$ on $(x^1, \ldots, x^{j - 1}, y, x^{j+1}, \ldots,  x^k)$.  Thus the conditional probability that $\mathbf{x}^j = y$ is
\begin{align*} 
\displaystyle \frac{  \mu^{\otimes k}(x^1, \ldots, x^{j - 1}, y, x^{j+1}, \ldots,  x^k)     }{ \sum_{z \text{ extends }u^j_t}\mu^{\otimes k}(x^1, \ldots, x^{j - 1}, z, x^{j+1}, \ldots,  x^k)    }  \   &=  \\
 \frac{  \mu(y) \cdot \prod_{j' \neq j}\mu(x^{j'})   }{ \sum_{z \text{ extends }u^j_t}\mu(z) \cdot \prod_{j' \neq j}\mu(x^{j'})   }  \ &= \\
 \frac{  \mu(y)    }{ \sum_{z \text{ extends }u^j_t}\mu(z)  }
\ &= \ \mu^{(u^j_t)}(y),
\end{align*}
by definition of $\mu^{(u^j_t)}$.  This proves Lemma~\ref{indlem}.
\end{proof}

Next, define collections 
\[\mathcal{X}  \ = \  \{X_{j, t}\}_{j \in [k], 0 \leq t \leq M} , \quad{} \mathcal{P} \ = \ \{P_t\}_{0 \leq t \leq M}  \]
of random variables, as follows. All the random variables are determined by the execution of $\mathcal{D}$ on an input drawn from $\mu^{\otimes k}$.  
Let $X_{j, t} :=W(u_j^t)$ if $|u_j^t| \leq T$; otherwise let $X_{j, t} :=1/2$.
Let $P_t := \prod_{j \in [k]}X_{j, t}$.

We claim that for each $0 \leq t < M$, $\bE[P_{t+1}] \leq \bE[P_{t}]$.  To see this, condition on any outcomes to the first $t$ queries, described by $u^1_t, \ldots, u^k_t$.  Now suppose that for the $(t + 1)$-st query, $\mathcal{D}$ queries the $i$-th bit of the $j$-th input ($i, j$ are determined by $u^1_t, \ldots, u^k_t$, since $\mathcal{D}$ is deterministic).  We note that $X_{j', t+1} = X_{j', t}$ for all $j' \neq j$.  If $|u_{j}^t| \geq T$ then also $X_{j, t+1} \leq X_{j, t}$, which implies $P_{t+1} \leq P_t$.  So assume $|u_{j}^t| < T$.  Then we have
\[\bE[P_{t+1}|u^1_t, \ldots, u^k_t] \ = \ \bE[X_{j, t+1} \cdot \prod_{j' \neq j}X_{j', t+1} | u^1_t, \ldots, u^k_t]  \]
\[= \ \bE[X_{j, t+1} | u^1_t, \ldots, u^k_t ] \cdot \prod_{j' \neq j}X_{j', t} \ \leq \ X_{j, t}\cdot \prod_{j' \neq j}X_{j', t} \ = \ P_t,  \]
 where we used Lemma~\ref{indlem} and part 2 of Lemma~\ref{Wlem}.  We conclude 
\[\bE[P_{t+1}] \ =  \ \bE[\bE[P_{t+1}| u^1_t, \ldots, u^k_t]] \ \leq \ \bE[P_t],\]
 as claimed. 
It follows that $\bE[P_M] \leq \bE[P_0]$.  
But we can bound $P_0$ directly: $P_0 = W(*^n)^k \leq (1 - \eps)^k$ (Lemma~\ref{Wlem}, part 1).  Thus $\bE[P_M] \leq (1 - \eps)^k$.

Now we argue that this implies an upper bound on the success probability of $\mathcal{D}$.   Condition on the bits $u^1_M, \ldots, u^k_M$ seen by $\mathcal{D}$ during a complete execution; these determine the $k$ output bits of $\mathcal{D}$.
For each $j \in [k]$, at least one of two possibilities holds: either $|u^j_M| > T$, or the $j$-th input is in a final state $\mu^{(u^j_M)}$ for which $\Pr_{\mathbf{y} \sim \mu^{(u_M^j)}}[f(\mathbf{y}) = 1] \in [1 - X_{j, M}, X_{j, M}]$.  Since the $k$ inputs remain independent under our conditioning, the conditional probability that $\mathcal{D}$ computes $f^{\otimes k}$ correctly is at most~$\prod_{j: |u^j_M| \leq T}X_{j, M}. $

$\mathcal{D}$ makes at most $\alpha \eps T k$ queries, so simple counting tells us that there are fewer than $\alpha \eps k$ indices $j$ for which $|u^j_M| > T$.  Thus,
\[  \displaystyle \prod_{j: |u^j_M| \leq T}X_{j, M} \ \leq \ \frac{\prod_{j \in [k]}X_j }{(\min_{j \in [k]} X_{j, M})^{\alpha \eps k}} 
\ \leq \  2^{\alpha \eps k} P_M \]
(since $X_{j, M} \geq 1/2$ for all $j$).  Taking expectations, we find that the \emph{overall} success probability of $\mathcal{D}$ is at most $\bE[ 2^{\alpha \eps k} P_M ] \leq  (2^{\alpha \eps}(1 - \eps))^k$.  

Finally, we simplify our bound.  We claim $2^x < 1 + .84x$ on $(0, 1/2]$.  To see this, just note that $2^0 = 1$, that $2^{1/2} < 1.42 = 1 + .84(1/2)$, and that $2^x$ is a convex function on $\bR$.  Then, since $0 < \alpha \eps \leq 1/2$, we have $2^{\alpha \eps}(1 - \eps) < (1+ .84\alpha \eps)(1 -  \eps) < 1 - \eps + .84\alpha\eps$.  The proof is complete.
\end{proof}

We remark that, as claimed in the Introduction, the proof above can be easily adapted to give an alternate proof of Shaltiel's optimal direct product theorem for ``fair'' algorithms making $Tk$ queries: we define the random variables $X_{j, t}$ exactly as before and note that $|u_{t}^j| \leq T$ for all $j, t$.

\section{Tightness of the bounds in Theorem~\ref{mainsdpt}}~\label{examplesec}

In this section we describe a family of functions and input distributions, due to \cite{Sha}, and explain why they show that the query/success tradeoff in Theorem~\ref{mainsdpt} is nearly best-possible, at least when $\alpha < .01$ and when $(1 - \eps)^k$ is also at most a small constant.

Fixing an integer $T > 0$, define $f_T: \{0, 1\}^{T + 2} \rightarrow \{0, 1\}$ as follows: let $f_T(x) := x_2$ if $x_1 = 1$, otherwise $f_T(x) := x_2 \oplus \ldots \oplus x_{T+2}$.  Given $\eps \in (0, 1/2)$, let $\mu_{\eps}$ be the distribution over $\{0, 1\}^{T +2}$ in which all bits are independent, $\Pr[x_1  = 1] = 1 - 2\eps$, and $\Pr[x_i = 1] = 1/2$ for all $i \in \{2, \ldots, T+2\}$.  Note that  if $\mathbf{y} \sim \mu_{\eps}$, a $T$-query-bounded algorithm can gain no information about the value of $f$ when $x_1 = 0$, so any such algorithm succeeds with probability at most $(1 - 2 \eps)1 + (2\eps)\frac{1}{2} = 1 - \eps$ in computing $f(\mathbf{y})$.  

Now consider the following algorithm $\mathcal{D}$ attempting to compute $f^{\otimes k}$ on inputs $(\mathbf{x}^1, \ldots, \mathbf{x}^k) \sim \mu_{\eps}^{\otimes k}$.  First $\mathcal{D}$ queries the first two bits of each input.  Call an input $\mathbf{x}^k$ ``bad'' if its first bit is 0, ``good'' if its first bit is 1.  Let $B \subseteq [k]$ denote the set of bad inputs.  Note that $\mathcal{D}$ learns the value of $f$ on each good input.  Next, $\mathcal{D}$ chooses arbitrarily a set $S \subseteq B$ of $\lfloor \alpha \eps k\rfloor$ bad inputs, and spends $T$ additional queries on each input in $S$ to determine the value of $f$ on these inputs (if there are fewer than $\lfloor \alpha \eps k\rfloor$ bad inputs, $\mathcal{D}$ queries them all and determines the value of $f^{\otimes k}$ with certainty).  Finally, $\mathcal{D}$ outputs the answer bits it has learned and makes random guesses for the remaining values.  

Observe that $\mathcal{D}$ uses at most $2k + \alpha \eps T k$ queries overall.
To analyze the success probability of $\mathcal{D}$, first consider an algorithm $\mathcal{D}'$ which uses only $2k$ queries to  look at the two bits of each input; $\mathcal{D}'$ outputs the correct value on good inputs, and guesses randomly on bad inputs.  It is easy to see that $\mathcal{D}'$ succeeds with probability $(1 - \eps)^k$ in computing $f^{\otimes k}$. Also, if $\mathcal{D}$ and $\mathcal{D}'$ are both run on a common $k$-tuple of inputs drawn from $\mu_{\eps}^{\otimes k}$, and we condition on the event that $|B| \geq \lfloor \alpha \eps k\rfloor$, then the success probability of $\mathcal{D}$ is $2^{\lfloor \alpha \eps k \rfloor}$ times the success probability of $\mathcal{D}'$, since the inputs are independent and $\mathcal{D}$ has $\lfloor \alpha \eps k \rfloor$ fewer random guesses to make.  Thus, $\Pr \left[\mathcal{D} \text{ succeeds} \right]$ is at least
\allowdisplaybreaks{\begin{align}  \label{happyeq}
&\Pr \left[|B| \geq \alpha \eps k \right] \cdot 2^{\lfloor \alpha \eps k \rfloor} \Pr \left[\mathcal{D}' \text{ succeeds}\bigg{|} |B| \geq \alpha \eps k \right] \nonumber  \\
&= \  2^{\lfloor \alpha \eps k \rfloor} \Pr \left[\mathcal{D}' \text{ succeeds} \wedge  |B| \geq \alpha \eps k \right] \nonumber   \\
&\geq \ 2^{\lfloor \alpha \eps k \rfloor}\cdot   \left(    \Pr \left[\mathcal{D}' \text{ succeeds}\right] -   \Pr \left[|B| < \alpha \eps k \right] \right)  \nonumber \\
&= \ 2^{\lfloor \alpha \eps k \rfloor}\cdot   \left(   (1 - \eps)^k -   \Pr \left[|B| < \alpha \eps k \right] \right).
\end{align}}
Define the indicator variable $Y_j := \mathbf{1}_{[j \notin B]}$; then the $Y_j$'s are independent, with $p = \Pr[Y_j = 1] = 1 - 2 \eps$.  Let $Y := Y_1 + \ldots + Y_k$. We apply Lemma~\ref{smallchern} to $Y$, with the settings $\delta := 2 \eps$ and $\beta := \alpha/2 \leq 1/2$, to obtain
\begin{align*}  
\Pr[|B| < \alpha \eps k] \ &= \ \Pr[Y > (1 - \alpha \eps)k] \\
&= \ \Pr[Y > (1 - (2\eps)(\alpha/2))k]\\
&< \ \left[ 1 - 2\eps + 6 (\alpha/2) \ln(2/\alpha) (2\eps)  \right]^{k}.   
\end{align*}
 This can be made less than $(1 - 1.5\eps)^{ k}$ if $\alpha$ is a small enough positive constant ($\alpha < .01$ will work).

Now if $(1 - \eps)^k$ is also at most a sufficiently small constant, then $(1 - 1.5\eps)^{k} < .1(1 - \eps)^k$ so that, by Eq.~\ref{happyeq},
\[  \Pr \left[\mathcal{D} \text{ succeeds} \right] \ > \ .9 \cdot 2^{\lfloor \alpha \eps k \rfloor} (1 - \eps)^k, \]
which is close to the maximum success probability allowed by Theorem~\ref{mainsdpt} if $\mathcal{D}$ used $\alpha \eps Tk$ queries. (Recall, though, that $\mathcal{D}$ uses $2k + \alpha \eps T k$ queries.)

\section{Proof of Theorem~\ref{worstcasesdpt}}\label{worstcasesec}

We now prove Theorem~\ref{worstcasesdpt} from the Introduction, our DPT for worst-case error, by combining Theorem~\ref{mainsdpt} with a version of Yao's minimax principle \cite{Yao}, which allows us to convert worst-case hardness assumptions in query complexity into average-case assumptions.

Define $R_{2, \delta}(f)$ as the minimum $T$ for which there exists a randomized $T$-query algorithm which computes $f(x)$ correctly with probability at least $1 - \delta$ for every $x$.  The following is a common version of Yao's principle, and can be proved directly using the minimax theorem of game theory.

\begin{lemma}\label{twosided_minimax} Fix $0 < \delta < 1/2$ and a Boolean function $f$.  There exists a distribution $\mu_{\delta}$ over inputs to $f$, such that every randomized algorithm making fewer than $R_{2, \delta}(f)$ queries succeeds in computing $f$ on $\mu_{\delta}$ with probability less than $1 - \delta$.
\end{lemma}

%XXX was namedproof
\begin{proof}[Proof of Theorem~\ref{worstcasesdpt}]  Let $f$ be given.  Let $\delta := 1/2 - \gamma/2$, and let $\mu := \mu_{\delta}$ be as provided by Lemma~\ref{twosided_minimax}.  Now fix a tiny constant $c \in (0, 1)$, and let $T := R_{2, \delta}(f) - c$; we have 
\[  \suc{T}{\mu}(f) \ \leq \ 1 - \eps, \]
for some value $\eps > \delta > 3/8$ (independent of $c$).
Now set $\alpha := \gamma$, and apply Theorem~\ref{mainsdpt} to find
\[\suc{\gamma \eps T k}{\mu}(f)  \ < \ \left(1 -  (1 - .84  \gamma)\eps \right)^k \ < \ \left(1 -  (1 - .84  \gamma)\delta \right)^k.   \]
Note that $\gamma \eps T k > \gamma \delta R_{2, \delta}(f) k $, if $c$ is chosen sufficiently small.
We conclude that any algorithm making at most $\gamma \delta R_{2, \delta}(f) k$ queries succeeds with probability less than
\begin{align*}
\left(1 -  (1 - .84  \gamma)\delta \right)^k \ &= \ \left(1 -  (1 - .84  \gamma)(1/2 - \gamma/2) \right)^k \\
&< \ \left(1/2 +  .42 \gamma +  \gamma/2 \right)^k \  <  \ \left(1/2 +  \gamma \right)^k
\end{align*}
in computing $f^{\otimes k}$ on inputs $\mathbf{x}^1, \ldots, \mathbf{x}^k \sim \mu^{\otimes k}$.  So, the worst-case success probability is also less than this amount.

Now we relate $ R_{2, \delta}(f)$ to $R_2(f)$ by standard sampling ideas.  Say $\mathcal{R}_{\delta}$ is an algorithm making $R_{2, \delta}(f)$ queries, which computes $f(x)$ with probability at least $1 - \delta = 1/2 + \gamma/2$ on each input.  Let $\mathcal{R}$ be the algorithm which given an input $x$, runs $\mathcal{R}_{\delta}(x)$ for $m := \lceil 3/\gamma^2 \rceil$ trials, outputting the majority value.  For $i \in [m]$, define the indicator variable $Y_i$ for the event $[\mathcal{R}_{\delta}$ succeeds on the $i$-th trial$]$, and let $Y := Y_1 + \ldots + Y_m$.  Then the probability that $\mathcal{R}(x)$ outputs an incorrect value is at most the probability that $Y \leq \bE[Y] - \gamma m / 2$, which by Hoeffding's inequality is at most $e^{ - 2 \gamma^2 m /4} \leq e^{-3/2} < 1/3$.  

Thus, $R_2(f) \leq   R_{2, \delta}(f)\cdot \lceil 3/\gamma^2 \rceil  <  4 R_{2, \delta}(f)/\gamma^2$ (using $\gamma  < 1/4$).  Then, we have
 \[\gamma^3 R_2(f) k/11 \ < \ \gamma (3/8)(\gamma^2 R_2(f)/4) k \  <  \ \gamma \delta R_{2, \delta}(f) k,\] 
from which Theorem~\ref{worstcasesdpt} follows.
\end{proof}

\section{Threshold direct product theorems}\label{gentsec}

In this section we prove our ``gambling lemma,'' Lemma~\ref{gamblers}, and use it to prove generalized threshold DPTs for relations (relation problems are formally defined in Section~\ref{tdsptsubsec}).  This will yield DPTs for non-Boolean functions as well as for the query complexity of learning tasks.  Further applications of Lemma~\ref{gamblers} will appear in later sections.

Let $\mathcal{P}([k])$ denote the collection of subsets of $[k]$.  Say that a subcollection $\mathcal{A} \subseteq \mathcal{P}([k])$ is \emph{monotone} if $[A \in \mathcal{A}, \ A \subseteq A']$ implies $A' \in \mathcal{A}$.  Monotone collections play an important role in what follows.

\subsection{A gambling lemma}\label{gamblersec}

Like the proof of Theorem~\ref{mainsdpt}, the statement of our next lemma is best explained by a gambling metaphor.  Suppose that a gambler gambles at $k$ tables, bringing an initial endowment of $p_j \in [0, 1]$ to the $j$-th table.  He cannot transfer funds between tables, or go into debt at any table; he can only play games for which his expected winnings are nonpositive; and the different tables' games use independent randomness.  However, the gambler can choose which game to play next at each table.

The gambler wants to reach a fortune of 1 at ``sufficiently many'' of the tables, where the meaning of ``sufficiently many'' is specified by a monotone subset $\mathcal{A} \subseteq \mathcal{P}([k])$.  One way the gambler may attempt to reach this goal is to simply place an ``all-or-nothing'' bet independently at each table; that is, at the $j$-th table, the gambler wins a fortune of 1 with probability $p_j$, and loses his $j$-th endowment with the remaining probability.  The following lemma states that this is in fact the gambler's best strategy.

\begin{lemma}\label{gamblers}  Suppose $k, N \geq 1$ are given, along with a collection $\{\mathcal{X}, \mathcal{U}\}$ of random variables (over a finite probability space).  Here $\mathcal{X} = \{\mathcal{X}_1, \ldots, \mathcal{X}_k \}$, where for each $j \in [k]$, $\mathcal{X}_j = \{X_{j, 0}, X_{j, 1}, \ldots, X_{j, N}\}$ is a sequence of variables in the range $[0, 1]$ \emph{(think of $X_{j, t}$ as the gambler's fortune at the $j$-th table after the first $t$ steps)}.  $\mathcal{U} = \{U_0, U_1, \ldots, U_{N - 1}\}$ is a sequence of random variables taking values over some finite set \emph{(think of $U_{t}$ as describing the form and outcomes of all gambles in the first $t$ steps)}.
Assume that for all $0\leq t < N$, $U_{t}$ determines $\{X_{1, t}, \ldots, X_{k, t}\}$, and also determines $U_{t'}$ for all $t' < t$.
Also assume that $\{X_{1, t+1}, \ldots, X_{k, t+1}\}$ are independent conditioned on $U_t$.
Then, if $X_{j, 0} \leq p_j \in [0, 1]$ for all $j \in [k]$, and $\mathcal{A}$ is a monotone subset of $\mathcal{P}([k])$, we have
\[\Pr[\{ j \in [k]: X_{j, N} = 1 \} \in \mathcal{A}] \ \leq \  \Pr[D \in \mathcal{A}],\]
where $D \subseteq [k]$ is generated by independently including each $j \in [k]$ in $D$ with probability $p_j$.
\end{lemma}

Note that we assume the gambler never attains a fortune greater than 1 at any table; this restriction is easily removed, but it holds naturally in the settings where we'll apply the Lemma.

\begin{proof}  We use the term ``$\mathcal{A}$-success'' to refer to the event $[\{ j \in [k]: X_{j, N} = 1\} \in \mathcal{A}]$ whose probability we are bounding.

We first make a simplifying observation: we claim it is without loss of generality to assume that between each consecutive times $(t, t+1)$, at most one of the fortunes changes, and that the fortune subject to change is determined by $t$.  Call a family of sequences with this property ``nice.''  To see this, consider any family $\mathcal{X}$ obeying Lemma~\ref{gamblers}'s assumptions, and modify it by ``splitting'' each transition $(t, t + 1)$ into a sequence of $k$ transitions, in the $j$-th of which the $j$-th fortune changes (according to the same distribution governing its transition in the original sequence).  

More formally, we define $\mathcal{X}'_j = \{X'_{j, 0}, \ldots, X'_{j, Nk}\}$ by letting $X'_{j, \ell} := X_{j, \lfloor   (\ell + k - j)/k \rfloor }$; and we define $\mathcal{U}' = \{U'_0, U'_1, \ldots, U'_{Nk - 1}\}$ by 
\[U'_{\ell} \ :=\  \left(U_{\lfloor \ell / k \rfloor}, \left(X'_{j, \ell'} \right)_{j \in [k], \ell' \leq \ell}   \right). \]  (We add extra information into $U'_{\ell}$ to ensure that it determines the random variables it is supposed to.) Lemma~\ref{gamblers}'s assumptions continue to hold for this modified, nice family of random variables; here we are using our original assumption that $\{X_{1, t+1}, \ldots, X_{k, t+1}\}$ are independent conditioned on $U_t$.  Also, the probability of $\mathcal{A}$-success is unchanged.  So let us assume from now on that $(\mathcal{X}, \mathcal{U})$ is nice, and for $0 \leq t < N$, let $j_t \in [k]$ be the index of the fortune subject to change between times $t$ and $t + 1$.

Fix any $k \geq 1$; we prove the statement by induction on $N \geq 1$.  First suppose $N = 1$, and let $j_0$ be as defined above.  Let $S \subseteq [k] \setminus \{j_0\}$ be the set of indices $j \neq j_0$ for which $p_{j} = 1$.  First suppose $S \in \mathcal{A}$; then $\Pr[D \in \mathcal{A}] = 1$, since each $j \in S$ is included in $D$ with probability 1.  In this case the conclusion is trivially satisfied.  Next suppose $S \cup \{j_0\} \notin \mathcal{A}$.  In this case, $\Pr[\mathcal{A}$-success$] = 0$, and again the conclusion is trivially satisfied.  So suppose $S \notin \mathcal{A}$, $S \cup \{j_0\} \in \mathcal{A}$, and condition on any value $U_0 = u$.  Then $\mathcal{A}$-success occurs iff $X_{j_0, 1} = 1$.  By Markov's inequality, $\Pr[X_{j_0, 1} = 1| U_0 = u ] \leq  \bE[X_{j_0, 1}|U_0 = u] \leq X_{j_0, 0} \leq p_{j_0} = \Pr[D \in \mathcal{A}]$.  This proves the statement for $N = 1$.  

So let $N > 1$ and assume the statement proved for $\{1, \ldots, N - 1\}$; we prove it for $N$.  Condition on any value $U_0 = u$, and condition further on the value $X_{j_0, 1} = a \in [0, 1]$.  The equalities $X_{j, 1} = X_{j, 0} \leq p_{j}$ are forced for all $j \neq j_0$; the residual collection of random variables $\{X_{j, t}: j \in [k], 1 \leq t \leq N\} \cup \{U_t: 1 \leq t < N\}$ under our conditioning obey Lemma~\ref{gamblers}'s assumptions, along with our added assumption; and these sequences are shorter by a step than our initial sequences.  Thus our induction hypothesis implies that 
\begin{equation}\label{gambeq1}\Pr[\mathcal{A}\text{-success}|U_0 = u, X_{j_0, 1} = a] \ \leq \ \Pr[D^{(a)} \in \mathcal{A}],\end{equation}
where $D^{(a)}$ is generated just like $D$ except that $j_0$ is now included in $D^{(a)}$ with probability $a$.

Let $q_0 := \Pr[D \setminus \{j_0\} \in \mathcal{A}]$ and $q_1 := \Pr[D \cup \{j_0\} \in \mathcal{A}]$.  Note that $q_0 \leq q_1$, since $\mathcal{A}$ is monotone.  We have 
\[\Pr[D^{(a)} \in \mathcal{A}] \ = \  (1 -  a)q_0   +  a q_1.\]
Taking expectations over $a$ in Eq.~\ref{gambeq1}, $\Pr[\mathcal{A}\text{-success}| U_0 = u]$ is at most
\begin{align*} 
&(1 - \bE[X_{j_0, 1}| U_0 = u])q_0  +   \bE[X_{j_0, 1}|U_0 = u] \cdot q_1 \\
&\leq  \  (1 - p_{j_0})q_0  + p_{j_0} q_1 \\
&\text{(since $q_0 \leq q_1$ and $\bE[X_{j_0, 1}|U_0 = u] \leq X_{j_0, 0} \leq p_{j_0}$)}\\
&= \ \Pr[D \in \mathcal{A}].
\end{align*}
As $u$ was arbitrary, this extends the induction to $N$, and completes the proof.
\end{proof}

\subsection{Application to threshold DPTs}\label{tdsptsubsec}

Now we prove our generalized threshold direct product theorem.  Our theorem will be within the framework of solving relation problems, a more general task than computing functions.  
A \emph{relation} (with Boolean domain) is a subset $P \subseteq \{0, 1\}^n \times B$, for some finite set $B$.  The relation is \emph{total} if for all $x \in \{0, 1\}^n$, there exists $b \in B$ such that $(x, b) \in P$.  For each total relation $P$ there is a natural computational problem: given an input $x$, try to output a $b$ for which $(x, b) \in P$.  Computing a function $f: \{0, 1\}^n \rightarrow B$ is equivalent to solving the relation problem for the total relation $P_f := \{(x, b): f(x) = b\}$.

If $\mathcal{R}$ is a (possibly randomized) query algorithm producing outputs in $B$, $P$ is a total relation, and $\mu$ a distribution, say that $\mathcal{R}$ \emph{$\eps$-solves $P$ with respect to $\mu$} if $\Pr_{x \sim \mu}[(x, \mathcal{R}(x)) \in P] \geq 1 - \eps$.
Define $\suc{T}{\mu}^{\text{rel}}(P) := 1 - \eps$, where $\eps \geq 0$ is the minimum value for which some $T$-query randomized algorithm $\mathcal{R}$ $\eps$-solves $P$ with respect to $\mu$.  As usual, this minimum exists and is attained by a deterministic height-$T$ decision tree.
For a randomized algorithm $\mathcal{R}$ making queries to $k \geq 1$ inputs $x = (x^1, \ldots, x^k)$ to $P$ and producing an output in $B^k$, let $\mathcal{R}_j(x) \in B$ be the $j$-th value outputted by $\mathcal{R}$.

Given $A, A' \subseteq [k]$, define the \emph{distance} $d(A, A') := |(A \setminus A') \cup (A' \setminus A) |$.  Given a set family $\mathcal{A} \subseteq \mathcal{P}([k])$, and a real number $r > 0$, define the \emph{strict $r$-neighborhood of $\mathcal{A}$}, denoted $N_r (\mathcal{A})$, as
\[  N_r (\mathcal{A})  \ :=   \  \{A': d(A, A') <  r \text{ for some }A \in \mathcal{A} \}.  \]
We have $\mathcal{A} \subseteq  N_r(\mathcal{A})$.  Note also that if $\mathcal{A}$ is monotone then so is $N_r(\mathcal{A})$.  We can now state our generalized threshold DPT:

\begin{theorem}\label{gen_tsdpt}  Fix a finite set $B$, and let $P \subseteq \{0, 1\}^n \times B$ be a total relation  for which \rm{}$\suc{T}{\mu}^{\text{rel}}(P) \leq 1 - \eps$\it{}.  Fixing any randomized algorithm $\mathcal{R}$ making queries to inputs $\mathbf{x} = (\mathbf{x}^1, \ldots , \mathbf{x}^k) \sim \mu^{\otimes k}$  and producing output in $B^k$, define the (random) set
\[S[\mathbf{x}] \ := \ \{j \in [k]: (\mathbf{x}^j, \mathcal{R}_j(\mathbf{x})) \in P\}.\]
Suppose $\mathcal{R}$ is $\alpha \eps Tk$-query-bounded for some $\alpha \in (0, 1]$, and $\mathcal{A}$ is any monotone subset of $\mathcal{P}([k])$.  Then:
\begin{enumerate}
\item[1.]\label{gent1} $\Pr[S[\mathbf{x}] \in \mathcal{A}]  \leq  |B|^{\alpha \eps k} \cdot \Pr[D \in \mathcal{A}]$,
where $D \subseteq [k]$ is generated by independently including each $j \in [k]$ in $D$ with probability $1 - \eps$.
\item[2.]\label{gent2} Also, for $D$ as above, $\Pr[S[\mathbf{x}] \in \mathcal{A}] \leq \Pr[D \in N_{\alpha \eps k}(\mathcal{A})]$.
\end{enumerate}
\end{theorem}

\begin{proof}  As in Theorem~\ref{mainsdpt}, we may assume $\eps, T > 0$, $\supp (\mu) = \{0, 1\}^n$.  We have $\eps \leq 1 - |B|^{-1}< 1$, since $P$ is total and an algorithm may output a random element of $B$.

For $u \in \{0, 1, *\}^n$ with $|u| \leq T$, let
\[   W_P(u)  \ :=  \  \suc{T - |u|}{\mu^{(u)}}^{\text{rel}}(P).     \]
Then $W_P(u) \in [|B|^{-1}, 1]$.  We have the following claim, whose  proof follows that of Lemma~\ref{Wlem}:

\begin{lemma} \label{gentWlem} 1. $W_P(*^n) \leq 1 - \eps$.  

\vspace{.5 em}

2. For any $u \in \{0, 1, *\}^n$ with $|u| < T$, and any $i \in [n]$, $\bE_{\mathbf{y} \sim \mu^{(u)}}[ W_P(u[x_i \leftarrow \mathbf{y}_i])  ] \leq W_P(u)$. 
\end{lemma}

Let $\mathcal{R}$ be $\alpha \eps T k$-query-bounded; as in Theorem~\ref{mainsdpt}, we may assume $\mathcal{R}$ is deterministic, so call it $\mathcal{D}$ instead.  Let $M := \lfloor \alpha \eps T k \rfloor$ as before, and recall the random strings $u^j_t$ defined in Theorem~\ref{mainsdpt}. 

Define random variables $\{X_{j, t}\}_{j \in [k], 0 \leq t \leq M}$, determined by an execution of $\mathcal{D}$ on inputs $(\mathbf{x}^1, \ldots , \mathbf{x}^k) \sim \mu^{\otimes k}$, by letting $X_{j, t} := W_P(u^j_t)$ if $|u^j_t| \leq T$, otherwise $X_{j, t} := |B|^{-1}$. 
Next, the natural idea is to apply Lemma~\ref{gamblers}.  First, however, we need to extend the sequences for one additional (non-query) step.  That is, we will define random variables $X_{j, M+1}$ for each $j \in [k]$.  We will use $\mathcal{X}$ to denote the collection of enlarged sequences.

Our definition of $X_{j, M + 1}$ depends on whether $|u^j_M| \leq T$, that is, on whether $\mathcal{D}$ made at most $T$ queries to $\mathbf{x}^j$ on the current execution.  If $|u^j_M| \leq T$, let $X_{j, M + 1} := \mathbf{1}_{[(\mathbf{x}^j, \mathcal{D}_j(\mathbf{x}) ) \in P]}$ be the indicator variable for the event that $\mathcal{D}$ solves $P$ on the $j$-th input.  If $|u^j_M| > T$, let $X_{j, M + 1}:= 1$ with probability $|B|^{-1}$, and let $X_{j, M + 1}:= 0$ with the remaining probability.  We let each such ``coin-flip'' be independent of the others and of $(\mathbf{x}^1, \ldots, \mathbf{x}^k)$.

Define the collection $\mathcal{U} = \{U_0, \ldots, U_M\}$ by $U_t := (u^1_t, \ldots, u^k_t)$.  We argue that the conditions of Lemma~\ref{gamblers} are satisfied by $(\mathcal{X}, \mathcal{U})$, with $N := M +1$.  First, for $0 \leq t' \leq t \leq M$, the stated conditions follow from Lemma~\ref{indlem} and part 2 of Lemma~\ref{gentWlem}.
Now consider the final, added step.  Condition on any value of $U_M = (u^1_M, \ldots, u^k_M)$.  Lemma~\ref{indlem} tells us that $\mathbf{x}^1, \ldots, \mathbf{x}^k$ are independent under this conditioning, and $\mathcal{D}$'s outputs are determined by $U_M$, so the variables $\{X_{j, M+1}\}$ are independent conditioned on $U_M$.  If $|u^j_M| \leq T$ then $\bE[X_{j, M+1} | U_M] \leq X_{j, M}$ by part 2 of Lemma~\ref{gentWlem}.  If $|u^j_M| > T$ then $\bE[X_{j, M+1}] = |B|^{-1} = X_{j, M}$.

Thus the assumptions of Lemma~\ref{gamblers} are satisfied, with $p_j = X_{j, 0} \leq 1 - \eps$.  We conclude that for any monotone $\mathcal{C} \subseteq \mathcal{P}([k])$, 
\begin{equation}\label{prev}\Pr[\{ j \in [k]: X_{j, N} = 1 \} \in \mathcal{C}] \ \leq \ \Pr[D \in \mathcal{C}],
\end{equation} 
where each $j \in [k]$ is independently included in $D$ with probability $1 - \eps$.

To prove statement 1 of Theorem~\ref{gen_tsdpt}, let $\mathcal{C} :=\mathcal{A}$.
Note that $S[\mathbf{x}]$ and $u^1_M, \ldots, u^k_M$ are determined by $\mathbf{x}$, since $\mathcal{D}$ is deterministic.  Condition on any value of $\mathbf{x}$ for which $S[\mathbf{x}] \in \mathcal{A}$.  Under this conditioning, if $j \in [k]$ satisfies $|u^j_M| \leq T$ and $j \in S[\mathbf{x}]$, then $X_{j, N} = 1$.  On the other hand, if $|u^j_M| > T$, then $[X_{j, N} = 1]$ holds with probability $|B|^{-1}$, and these events are independent for each such $j$.  By the query bound on $\mathcal{D}$, there are fewer than $\alpha \eps k$ indices $j$ in our conditioning for which $|u^j_M| > T$.  Thus, 
\[   \Pr[    \{ j \in [k]: X_{j, N} = 1 \} \in \mathcal{A}  | S[\mathbf{x}]  \in \mathcal{A} ]   \  \geq \  |B|^{-\alpha\eps k},  \]
 which in combination with Eq.~\ref{prev} implies
\[  \Pr[   S[\mathbf{x}]  \in \mathcal{A}]  \ \leq  \   |B|^{\alpha \eps k} \cdot  \Pr[D \in \mathcal{A}],   \]
as needed.  To prove statement 2 of Theorem~\ref{gen_tsdpt}, let $\mathcal{C} := N_{\alpha \eps k}(\mathcal{A})$ in Eq.~\ref{prev}: we find
\begin{equation*}\label{prevb}\Pr[\{ j \in [k]: X_{j, N} = 1 \} \in N_{\alpha \eps k}(\mathcal{A})] \   \leq \ \Pr[D \in N_{\alpha \eps k}(\mathcal{A})].
\end{equation*}
Arguing as above, $S[\mathbf{x}] \setminus \{ j \in [k]: X_{j, N} = 1 \}$ is always a set of size less than $\alpha \eps k$, so $[S[\mathbf{x}] \in \mathcal{A}]$ implies $[\{ j \in [k]: X_{j, N} = 1 \} \in N_{\alpha \eps k}(\mathcal{A})]$.  Thus, we have
$\Pr[ S[\mathbf{x}] \in \mathcal{A}] \leq \Pr[D \in N_{\alpha \eps k}(\mathcal{A})]$.
\end{proof}

Part 1 of Theorem~\ref{gen_tsdpt} is a proper generalization of Theorem~\ref{mainsdpt}.  To see this, just set $\mathcal{A} := \{[k]\}$, $P := P_f$, and note that in this case, $\Pr[D \in \mathcal{A}] = (1 - \eps)^k$.
As another dividend, we obtain the following threshold DPT for relations, which specializes to an ordinary DPT for this setting (statement 3 in the Theorem below).

\begin{theorem}\label{tsdpt} Let $P \subseteq \{0, 1\}^n \times B$ be a total relation for which \rm{}$\suc{T}{\mu}^{\text{rel}}(P) \leq 1 - \eps$\it{}.  Fix any $\eta \in (0, 1]$.  For any randomized algorithm $\mathcal{R}$ making queries to inputs $\mathbf{x} = (\mathbf{x}^1, \ldots , \mathbf{x}^k) \sim \mu^{\otimes k}$, define the (random) set $S[\mathbf{x}]$ as in Theorem~\ref{gen_tsdpt}.
Then if $\mathcal{R}$ is $\alpha \eps Tk$-query-bounded for $\alpha \in (0, 1]$, we have:
\begin{enumerate} 
\item[1.] $\Pr[|S[\mathbf{x}]| \geq \eta k]  \leq |B|^{\alpha \eps k} \cdot \Pr_{Y \sim B_{k, 1 - \eps}}[ Y \geq \eta k]$, and also
\item[2.] $\Pr[|S[\mathbf{x}]| \geq \eta k] \leq  \Pr_{Y \sim B_{k, 1 - \eps}}[ Y \geq (\eta - \alpha \eps) k]$.
\item[3.] $\Pr[|S[\mathbf{x}]|  =  [k]]$ is at most the minimum of $|B|^{\alpha \eps k}(1 - \eps)^k$ and $\Pr_{Y \sim B_{k, 1 - \eps}}[ Y \geq (1 - \alpha \eps)k]$.   If $\alpha \leq 1/2$ the second bound in the $\min$ is at most $ \left[   1 - \eps + 6\alpha \ln(1/\alpha) \eps   \right]^{k}$.   
\end{enumerate} 
\end{theorem}

\begin{proof} Apply parts 1 and 2 of Theorem~\ref{gen_tsdpt}, with the choice $\mathcal{A} := \{A \subseteq [k]: |A| \geq \eta k\}$.  We have $\Pr[D \in \mathcal{A}] = \Pr[ D_1 + \ldots + D_k \geq \eta k]$, where we define $D_j := \mathbf{1}_{[j \in D]}$.  These $0/1$-valued variables are independent with bias $1 - \eps$, which gives statement 1.  Similarly, $\Pr[D \in N_{\alpha \eps k}(\mathcal{A})] = \Pr[ D_1 + \ldots + D_k \geq (\eta - \alpha \eps) k]$, which gives statement 2.  Statement 3 simply combines statements 1 and 2, under the setting $\eta = 1$.  For the final bound in statement 3, we apply Lemma~\ref{smallchern} with $\beta := \alpha$, $\delta := \eps$.
\end{proof}

Theorem~\ref{special_tsdpt} in the Introduction follows from the special case of Theorem~\ref{tsdpt} in which $P := P_f$.

The success bound $|B|^{\alpha \eps k}(1 - \eps)^k$ appearing above can also be derived by an easy modification of the proof of Theorem~\ref{mainsdpt}, in which the condition $X_{j, t} \geq 1/2$ we exploit becomes $X_{j, t} \geq |B|^{-1}$.  When $|B|$ is large, however, the alternative bound provided in Theorem~\ref{tsdpt} will tend to give better results.

Note that part 2 of Theorem~\ref{tsdpt}, in conjunction with Chernoff inequalities, gives success bounds which decay exponentially in $k$ for any fixed $\alpha, \eps, \eta$ for which $\eta > 1 - \eps + \alpha \eps$.  Shaltiel's examples, described in Section~\ref{examplesec}, show that this cutoff is nearly tight: on those functions, the algorithm $\mathcal{D}$  described in Section~\ref{examplesec} makes $2k + \alpha \eps T k$ queries and (it is easily checked) typically solves about $(1 - \eps + .5\alpha \eps)k$ of the instances correctly.

Threshold DPTs for the worst-case setting can also be derived from Theorems~\ref{gen_tsdpt} and \ref{tsdpt}, by the same reduction to the average-case setting used to prove Theorem~\ref{worstcasesdpt}.

\subsection{Direct product theorems for learning tasks}

Theorems~\ref{gen_tsdpt} and \ref{tsdpt} readily imply direct product theorems for the query complexity of certain learning tasks, as we explain next.  Consider the scenario in which a randomized algorithm $\mathcal{R}$ is given query access to an unknown function $h: \{0, 1\}^n \rightarrow \{0, 1\}$ drawn from some distribution $\mu$ over a hypothesis class $\mathcal{H}$.  That is, for any string $x$, $\mathcal{R}$ can query the value $h(x)$.  The algorithm $\mathcal{R}$ attempts to output a hypothesis $\tilde{h}$ which is ``close'' to $h$.  That is, we fix some symmetric relation $\close \subseteq \mathcal{H} \times \mathcal{H}$ (assume $\close(h, h)$ always holds), and we wish to find some $\tilde{h}$ such that $\close(h, \tilde{h})$ holds.

This task can be equivalently modeled as the relation problem associated with the total relation 
\[P_{\mathcal{H}} \ := \ \{(h, h'): h, h' \in \mathcal{H} \wedge \close(h, h')\}, \]
where $h$ is given in truth-table form as a Boolean string, under the input distribution $h \sim \mu$.  (We don't give a membership criterion for $P_{\mathcal{H}}$ when $h \notin \mathcal{H}$; this is unimportant since $\supp(\mu) \subseteq \mathcal{H}$.)

In the $k$-fold learning problem associated with $\mathcal{H}, \mu$, the algorithm has query access to each of $k$ functions $(h_1, \ldots, h_k) \sim \mu^{\otimes k}$,
 and the goal is to output guesses $\tilde{h}_1, \ldots \tilde{h}_k$ such that $\close(h_j, \tilde{h}_j)$ holds for all (or at least ``many'') indices $j \in [k]$.  This task is equivalent to the $k$-fold relation problem associated with $P_{\mathcal{H} }$, and Theorems \ref{gen_tsdpt} and \ref{tsdpt} apply.

\section{Proof of the XOR lemma}\label{xorsec}

The proof of our XOR Lemma, Theorem~\ref{xorsdpt} from the Introduction, is modeled on the proof of our threshold DPTs, and reuses Lemma~\ref{gamblers}.  %XXX begin namedproof
\begin{proof}[Proof of Theorem~\ref{xorsdpt}]  As usual we first set up some preliminaries.  For a deterministic algorithm $\mathcal{D}$ over $n$ input bits define 
\[W_{\oplus}(u) \ := \ 2 \cdot \suc{T - |u|}{\mu^{(u)}}(f)  - 1.\]

\begin{lemma} \label{xorWlem} 1. $W_{\oplus}(*^n) \leq 1 - 2\eps$.

\vspace{.5 em}

2. For any $u \in \{0, 1, *\}^n$ with $|u| < T$, and any $i \in [n]$, $\bE_{\mathbf{y} \sim \mu^{(u)}}[ W_{\oplus}(u[x_i \leftarrow \mathbf{y}_i])  ] \leq W_{\oplus}(u)$. 
\end{lemma}

Lemma~\ref{xorWlem} follows immediately from Lemma~\ref{Wlem}, since $W_{\oplus}(u) = 2 W(u) - 1$.

Now we prove the Theorem.  As in the proof of Theorem~\ref{mainsdpt}, we may assume $\eps, T > 0$, $\supp (\mu) = \{0, 1\}^n$, and it is enough to prove the success bound for each deterministic $\alpha \eps T k$-query algorithm $\mathcal{D}$ attempting to solve $f^{\oplus k}(\mathbf{x}^1, \ldots, \mathbf{x}^k)$ on inputs $\mathbf{x}^1, \ldots, \mathbf{x}^k \sim \mu^{\otimes k}$.  
Recall the definitions of $u^j_t$ (for $j \in [k], 0 \leq t \leq M := \lfloor \alpha \eps Tk \rfloor$) from Theorem~\ref{mainsdpt}.  
For a deterministic algorithm $\mathcal{D}$  define $\{X_{j, t}\}_{j \in [k], 0 \leq t \leq M}$ as follows: if $|u^j_t| \leq T$, set $X_{j, t} := W_{\oplus}(u_j^t)$; otherwise, set $X_{j, t} := 0$.  

We will extend the random sequences $\{X_{j, t}\}$ for one additional (non-query) step, and will let $\mathcal{X}$ denote our enlarged collection.  To set up our extension, we first define random variables $b_j, r_j, a_j$ for $j \in [k]$, determined by $u^j_M$, as follows.  Let $b_j \in \{0, 1\}$ be defined as the likeliest value of $f(\mathbf{y})$, where $\mathbf{y} \sim \mu^{(u^j_M)}$ (break ties arbitrarily).  Let $r_j := \Pr[ f(\mathbf{y}) = b_j ] \in [1/2, 1]$, where again $\mathbf{y}\sim \mu^{(u^j_M)}$.  Let $a_j := 2r_j - 1 \in [0, 1]$.

If $|u^j_M| > T$, set $X_{j, M+1} := 0$.  If instead $|u^j_M| \leq T$, our random process ``inspects'' the actual value of the bit $f(\mathbf{x}^j)$ to help determine $X_{j, M + 1}$.  If $f(\mathbf{x}^j) \neq b_j$, let $X_{j, M+ 1} := 0$.  If $f(\mathbf{x}^j) = b_j$, let $X_{j, M + 1} := 1$ with probability $a_j/r_j$, and $X_{j, M + 1} := 0$ with the remaining probability, where this random decision is independent of all others.  Thus in this case, 
\[\bE[X_{j, M + 1} | u^1_M, \ldots, u^k_M] \ =  \ r_j \cdot (a_j/r_j)  \ = \ a_j \ \leq \ X_{j, M},\]
where the last inequality holds by the definition of $W_{\oplus}(u^j_M)$ since $|u^j_M| \leq T$.

Let $\mathcal{U} = (U_0, \ldots, U_{M})$, where $U_t := (u^1_t, \ldots, u^k_t)$.  By an argument analogous to that in the proof of Theorem~\ref{gen_tsdpt}, we verify that $(\mathcal{X}, \mathcal{U})$ obey the assumptions of Lemma~\ref{gamblers}, this time with $p_j := 1 - 2\eps$.  Applying Lemma~\ref{gamblers} to $\mathcal{A} := \{A \subseteq [k]: |A| > (1 - \alpha \eps)k \}$, we find
\begin{equation}\label{jollyeq}\Pr[|\{j: X_{j, M + 1} = 1  \}| > (1 - \alpha \eps)k ] \ \leq \  \Pr[D \in \mathcal{A}],
\end{equation}
where each $j \in [k]$ is independently included in $D$ with probability $(1 - 2\eps)$.  We have $\Pr[D \in \mathcal{A}] = \Pr_{Y \sim B_{k, 1 - 2\eps}}[Y > (1 - \alpha\eps)k]$.

We analyze events $F$ of form $F:= [U_M = (u^1_M, \ldots, u^k_M), X_{1, M+1} =z_1, \ldots, X_{k, M+1} = z_k]$. Note that conditioning on $F$ does \emph{not} condition on the particular values $f(\mathbf{x}^{j})$ which helped determine the values $z_j$.  Focus attention on any such event $F$ for which $|\{j: X_{j, M + 1} = 1  \}| \leq (1 - \alpha \eps)k$.    Since $\mathcal{D}$ makes at most $\alpha\eps T k$ queries, there are fewer than $\alpha\eps k$ indices $j$ for which $|u^j_M| > T$.  In particular, there exists a $j^{\star} \in [k]$ for which $|u^{j^{\star}}_M| \leq T$ \emph{and} $X_{j^{\star}, M + 1} < 1$ (so, by our definitions, $X_{j^{\star}, M+1} =0$). 

Now let the event $F'$ be defined just like $F$, except that $F'$ makes no conditioning on $X_{j^{\star}, M+1}$ (so, $F = F' \wedge [X_{j^{\star}, M+1} = 0]$).  Then,
\[\Pr[f(\mathbf{x}^{j^{\star}}) = b_{j^{\star}} | F] = \Pr[f(\mathbf{x}^{j^{\star}}) = b_{j^{\star}}| F' \wedge X_{j^{\star}, M+1} = 0]  \]
\[= \ \frac{\Pr[f(\mathbf{x}^{j^{\star}}) = b_{j^{\star}} \wedge X_{j^{\star}, M+1} = 0  | F']}{ \Pr[X_{j^{\star}, M+1} = 0   | F']  } \]
\[ = \ \frac{\Pr[f(\mathbf{x}^{j^{\star}}) = b_{j^{\star}}|F'] \cdot \Pr[ X_{j^{\star}, M+1} = 0  | F',f(\mathbf{x}^{j^{\star}}) = b_{j^{\star}} ]}     									{ \sum_{b \in \{0, 1\}} \Pr[f(\mathbf{x}^{j^{\star}}) = b|F'] \cdot \Pr[ X_{j^{\star}, M+1} = 0  | F',f(\mathbf{x}^{j^{\star}}) = b ]  }      \] 
\[   = \  \frac{      r_{j^{\star}} (1 - a_{j^{\star}}/r_{j^{\star}})        }{     r_{j^{\star}} (1 - a_{j^{\star}}/r_{j^{\star}})     +   (1 - r_{j^{\star}})\cdot 1     } \]
(using the fact that $\mathbf{x}^{1}, \ldots, \mathbf{x}^{k}$ are independent conditioned on $U_M$, by Lemma~\ref{indlem}, and the additional fact that $\{X_{j, M+1}\}_{j \in [k]}$ are independent conditioned on $U_M$)
\[ = \ \frac{r_{j^{\star}} - a_{j^{\star}} }{1 - a_{j^{\star}}} \ = \ \frac{\frac{1}{2}(1 + a_{j^{\star}}) -a_{j^{\star}} }{1 -   a_{j^{\star}}  } \ = \ 1/2 .    \] 
Thus, $f(\mathbf{x}^{j^{\star}})$ is an unbiased random bit conditioned on $F$.  Consequently, $f^{\oplus k}(  \mathbf{x}^{1}, \ldots, \mathbf{x}^{k}  ) = f(\mathbf{x}^{j^{\star}}) \oplus f^{\oplus k - 1}(\mathbf{x}^{1}, \ldots, \mathbf{x}^{j^{\star}-1}, \mathbf{x}^{j^{\star}+1}, \ldots,   \mathbf{x}^{k} )$ is an unbiased random bit conditioned on $F$.  Thus under this conditioning, $\mathcal{D}$'s output bit equals the $k$-fold XOR with probability exactly $1/2$.  Now $F$ was an arbitrary outcome of $U_M, X_{1, M+1}, \ldots, X_{k, M+1}$ for which $|\{j: X_{j, M + 1} = 1  \}| \leq (1 - \alpha \eps)k$.  It follows that
\begin{align*}
&\Pr_{\mathbf{x} \sim \mu^{\otimes k}}[\mathcal{D}(\mathbf{x}) = f^{\oplus k}(\mathbf{x})] \ \leq  \  \Pr\left[|\{j: X_{j, M + 1} = 1  \}| > (1 - \alpha \eps)k\right]  +  \\
 &\quad{} \quad{} \quad{} \quad{}\quad{} \quad{} \frac{1}{2}\Pr\left[|\{j: X_{j, M + 1} = 1  \}| \leq (1 - \alpha \eps)k\right]\\
&= \ \frac{1}{2}\left(   1  +   \Pr\left[|\{j: X_{j, M + 1} = 1  \}| > (1 - \alpha \eps)k\right]\right)\\
&\leq \  \frac{1}{2} \left(1 + \Pr_{Y \sim B_{k, 1 - 2\eps}}[Y > (1 - \alpha\eps)k]\right) ,
\end{align*}
using Eq.~\ref{jollyeq}.  

Finally, to get the concrete bound claimed in statement of Theorem~\ref{xorsdpt}, first suppose $\eps = 1/2$; in this case the bound follows easily since $Y = 0$ with certainty.  If $\eps < 1/2$,
note that $(1 - \alpha\eps)k = (1 - (\alpha/2)(2\eps))$, and apply Lemma~\ref{smallchern} with $\delta := 2\eps < 1$ and $\beta := \alpha/2 \leq 1/2$.
\end{proof}

\section{Direct product theorems for search problems and errorless heuristics}\label{searchsec}

We define a fairly general notion of search problems in the query model for which a direct product theorem can be proved.   We will also obtain a DPT for \emph{errorless heuristics}, defined in Section~\ref{errorless_subsec}.

\subsection{Search problems}\label{searchsubsec}

We need some preliminary definitions.  
Given $u, v \in \{0, 1, *\}^n$, say that $u$ and $v$ \emph{agree} if $u_i \in \{0, 1\}$ implies $v_i \in \{*, u_i\}$.  Note that this definition is symmetric in $u$ and $v$.  If $u, v$ agree, define their \emph{overlay} $u \circ v \in \{0, 1, *\}^n$ by $(u \circ v)_i := b \in \{0, 1\}$ if either $u_i = b$ or $v_i = b$, otherwise $(u \circ v)_i := *$.  Say that $u$ \emph{extends $v$} if $v_i \in \{0, 1\}$ implies $u_i = v_i$.

Say we are given a distribution $\mu$ on $\{0, 1\}^n$, and a (possibly randomized) query algorithm $\mathcal{R}$; if $\mathcal{R}$ runs on an input distributed according $\mu$, we denote by $U_{\mathcal{R}, \mu} \in \{0, 1, *\}^n$ the random string describing the input bits seen by $\mathcal{R}$.

A \emph{search problem} is defined by a subset $V \subseteq \{0, 1, *\}^n$.  We say that $\mathcal{R}$ \emph{$\eps$-solves} the search problem $V$ with respect to an input distribution $\mu$ over $\{0, 1\}^n$ if, with probability $\geq  1 - \eps$, $U_{\mathcal{R}, \mu}$ extends some $v \in V$.  (We allow the possibility that some $x \in \supp(\mu)$ do not extend \emph{any} $v \in V$.)
 Define $\suc{T}{\mu}(V) := 1 - \eps$, where $\eps$ is the minimal value such that some $T$-query randomized algorithm $\eps$-solves search problem $V$ on inputs from $\mu$.

Define the $k$-fold search problem $V^{\otimes k} := \{(v^1, \ldots, v^k): v^j \in V, \forall j \in [k]\} \subseteq \{0, 1, *\}^{kn}$.  Thus to solve $V^{\otimes k}$, an algorithm must solve each of the $k$ constituent search problems.  We generalize this notion in order to state a threshold DPT, which will imply our ordinary DPT.  For a monotone subset $\mathcal{A} \subseteq \mathcal{P}([k])$, define 
\[V^{k, \mathcal{A}}  \ := \  \{(v^1, \ldots, v^k  ):  \{j \in [k]: v^j \in V\} \in \mathcal{A}    \}.   \]
Thus to solve $V^{k, \mathcal{A}}$, an algorithm must solve ``sufficiently many'' of the $k$ search problems, as specified by $\mathcal{A}$.

Recall the notation $N_{r}(\cdot)$ from Section~\ref{gentsec}.  Our generalized threshold DPT for search problems is as follows:

\begin{theorem}\label{search_gen_tsdpt}  Suppose the search problem $V$ satisfies $\suc{T}{\mu}(V) \leq 1 - \eps$.  Then for any $\alpha \in (0, 1]$ and any monotone $\mathcal{A} \subseteq \mathcal{P}([k])$, 
\[\suc{\alpha \eps T k}{\mu^{\otimes k}}(V^{k, \mathcal{A}}) \ \leq \ \Pr[D \in N_{\alpha \eps k}(\mathcal{A})],    \]
where each $j \in [k]$ is independently included in $D$ with probability $1 - \eps$.
\end{theorem}

\begin{proof}  In the search setting, $\eps$ can potentially be any value in $[0, 1]$.  The boundary cases are trivial, so assume $0 < \eps < 1$.  As usual, we can assume that $T > 0$ and $\supp(\mu) = \{0, 1\}^n$, and it is enough to bound the success probability of any deterministic $\alpha \eps T k$-query algorithm.

Following Theorem~\ref{mainsdpt}, we first develop some concepts related to a computation on a single input to the search problem $V$.
For each $u \in \{0, 1, *\}^n$ for which $|u| \leq T$, let $\val_V(u) := 1$ if $u$ extends some $v \in V$, otherwise $\val_V(u) :=0$.  For a deterministic query algorithm $\mathcal{D}$ let $W_V(u, \mathcal{D}) := \bE[\val(u \circ U_{\mathcal{D}, \mu^{(u)}})]$.  (Note that $u$ and $U_{\mathcal{D}, \mu^{(u)}}$ always agree.)  

If $|u| \leq T$, let $W_{V}(u) := \max_{\mathcal{D}}(W_{V}(u, \mathcal{D}))$, where the maximum ranges over all deterministic algorithms making at most $T - |u|$ queries.  In other words, $W_{V}(u)$ is the maximum success probability of any $(T - |u|)$-query algorithm in solving $V$ on an input $\mathbf{y} \sim \mu^{(u)}$, where we reveal the bits described by $u$ ``for free'' to the algorithm.
Then we have:

\begin{lemma} \label{searchWlem} 1. $W_V(*^n) \leq 1 - \eps$.

\vspace{.5 em}
  
2. For any $u \in \{0, 1, *\}^n$ with $|u| < T$, and any $i \in [n]$, $\bE_{\mathbf{y} \sim \mu^{(u)}}[ W_V(u[x_i \leftarrow \mathbf{y}_i])  ] \leq W_V(u)$.  
\end{lemma}
We omit the proof, which is essentially the same as that of Lemma~\ref{Wlem}.

Let $\mathcal{D}$ be any deterministic algorithm making at most $M := \lfloor \alpha \eps T k \rfloor$ queries and attempting to compute $V^{k, \mathcal{A}}$ on inputs drawn as $(\mathbf{x}^1, \ldots, \mathbf{x}^k) \sim \mu^{\otimes k}$.  
For $0 \leq t \leq M$, and for $j \in [k]$, let $u_t^{j}$ be defined as in the previous proofs.  Let $\mathcal{X} = \{X_{j, t}\}_{j \in [k], 0 \leq t \leq M}$, where $X_{j, t} :=W_{V}(u_t^j)$ if $|u_t^j| \leq T$, otherwise $X_{j, t} := 0$.  

Unlike in Theorem~\ref{gen_tsdpt}, we have no need to add any additional steps to our random sequences.  For $0 \leq t < M$, we let $U_t := (u^1_t, \ldots, u^k_t)$ just as before.  Setting $N := M$ and reasoning as in Theorem~\ref{gen_tsdpt}, we verify that the assumptions of Lemma~\ref{gamblers} are satisfied, with $p_j = X_{j, 0} \leq 1 - \eps$ (Lemma~\ref{searchWlem}, part 1).

Applying Lemma~\ref{gamblers} to the monotone set $N_{\alpha \eps k}(\mathcal{A})$, we conclude that
\begin{equation}\label{prev2}\Pr[\{ j \in [k]: X_{j, M} = 1 \} \in N_{\alpha \eps k}(\mathcal{A})] \  \leq \  \Pr[D \in N_{\alpha \eps k}(\mathcal{A})],
\end{equation}
where each $j \in [k]$ is independently included in $D$ with probability $1 - \eps$.

Now condition on any execution of $\mathcal{D}$, and consider any $j \in [k]$ such that $X_{j, M} < 1$.  By our definitions, at least one of two possibilities holds: either $|u^j_M| > T$ (there are fewer than $\alpha \eps k$ such indices $j$), or $u^j_M$ does not extend any $v \in V$.  
Thus if $\mathcal{D}$ solves the search problem $V^{k, \mathcal{A}}$ on the present execution, we have $\{ j \in [k]: X_{j, M} = 1 \} \in N_{\alpha \eps k}(\mathcal{A})$.
Combining this with Eq.~\ref{prev2} yields the Theorem.
\end{proof}

From Theorem~\ref{search_gen_tsdpt}, we will directly obtain a standard threshold DPT and an ordinary DPT for search problems.  First, given a search problem $V \subseteq \{0, 1, *\}^n$ and a real number $s \in [0, k]$, define $\mathcal{C}[\geq s] := \{A \subseteq [k]: |A| \geq s\}$. 

\begin{theorem}\label{search_tsdpt} Suppose $\suc{T}{\mu}(V) \leq 1 - \eps$.  Then for any $\alpha \in (0, 1]$ and any $\eta \in (0, 1]$, 
\[\suc{\alpha \eps T k}{\mu^{\otimes k}}(V^{k, \mathcal{C}[\geq \eta k]}) \ \leq \ \Pr_{Y \sim B_{k, 1 - \eps}}\left[Y >    (\eta - \alpha \eps)k\right].    \]
\end{theorem}

\begin{proof}  Apply Theorem~\ref{search_gen_tsdpt} with $\mathcal{C} := \mathcal{C}[\geq \eta k]$, and note that $D \in N_{\alpha \eps k} \left(\mathcal{C}[\geq \eta k] \right)$ iff $ |D| > \eta k - \alpha \eps k$, which is equivalent to $\left[ D_1 + \ldots + D_k >    (\eta - \alpha \eps)k \right]$, where $D_j := \mathbf{1}_{[j \in D]}$.  These indicator variables are independent with expectation $1 - \eps$.
\end{proof}

\begin{theorem}\label{search_sdpt} Suppose $\suc{T}{\mu}(V) \leq 1 - \eps$.  Then for any $\alpha \in (0, 1]$, 
\[\suc{\alpha \eps T k}{\mu^{\otimes k}}(V^{\otimes k}) \ \leq \  \Pr_{Y \sim B_{k, 1 - \eps}}\left[Y >    (1 - \alpha \eps)k\right] .    \]
\end{theorem}

\begin{proof}  Note that $V^{\otimes k} = V^{k, \mathcal{C}[\geq k]}$, so the result follows from Theorem~\ref{search_tsdpt} with $\eta := 1$.
\end{proof}

\subsection{Errorless heuristics}\label{errorless_subsec}

An \emph{errorless heuristic} for a (not necessarily Boolean) function $f$ is a randomized query algorithm $\mathcal{R}$ outputting values in $\{0, 1, ?\}$ such that for all $x$, $\mathcal{R}(x) \in \{f(x), ?\}$ with probability 1.  We say that an errorless heuristic $\mathcal{R}$ \emph{$\eps$-solves $f$ with zero error} with respect to input distribution $\mu$ if $\Pr_{x \sim \mu}[\mathcal{R}(x) = f(x)] \geq 1 - \eps$.  Let $\suc{T}{\mu}^{\zerr}(f) := 1 - \eps$, where $\eps$ is the minimal value such that some $T$-query errorless heuristic $\eps$-solves $f$ with zero error with respect to $\mu$.  Note that $\suc{T}{\mu}^{\zerr}(f)$ is exactly $\suc{T}{\mu}(V_f)$, where the search problem $V_f$ is defined as 
\[V_f \ := \ \{u \in \{0, 1, *\}^n: u \text{ forces the value of } f\}.\] 
Also, note that $V_{f^{\otimes k}} = V_f^{\otimes k}$. Thus the following result is immediately implied by Theorem~\ref{search_sdpt}:

\begin{theorem}\label{zeroerrorsdpt}  Suppose $\suc{T}{\mu}^{\zerr}(f) \leq 1 - \eps$.  Then for $\alpha \in (0, 1]$, 
\[\suc{\alpha \eps T k}{\mu^{\otimes k}}^{\zerr}(f^{\otimes k}) \ \leq \ \Pr_{Y \sim B_{k, 1 - \eps}}\left[Y >    (1 - \alpha \eps)k\right]. \]  
\end{theorem}

Let us revisit the XOR problem in the current setting.  It is easy to see that an errorless heuristic to compute the $k$-fold XOR $f^{\oplus k}$, on inputs drawn from a product distribution, cannot produce any output other than ``$\ ?\ $'' unless its queries allow it to determine the value of $f^{\otimes k}$.  Thus Theorem~\ref{zeroerrorsdpt} also implies an XOR lemma with the same success bound for errorless heuristics.

Next we prove a worst-case analogue of Theorem~\ref{zeroerrorsdpt}.
Define $R_0(f)$, the \emph{zero-error randomized query complexity of $f$}, as the minimum $T$ for which some algorithm $\mathcal{R}$ outputs $f(x)$ with probability 1 for each $x$, and for which the \emph{expected} number of queries made by $\mathcal{R}$ to any input is at most $T$.  
The following is another variant of Yao's minimax principle \cite{Yao}; we include a proof for completeness.

\begin{lemma}\label{heurlem} Let $\eta \in (0, 1]$.  There exists a distribution $\mu_{\eta}$ over inputs to $f$, such that $\suc{\eta R_0(f)}{\mu_{\eta}}^{\zerr}(f) \leq \eta$.
\end{lemma}

\begin{proof}  Consider the following 2-player game: player 1 chooses a (possibly randomized) errorless heuristic $\mathcal{R}$ for $f$ which makes at most $\eta R_0(f)$ queries, and player 2 chooses (simultaneously) an input $x$ to $f$.  Player 1 wins if $\mathcal{R}(x) = f(x)$.  We claim there exists a randomized strategy for player 2, that is, a distribution $\mu =: \mu_{\eta}$ over inputs to $x$, that beats any strategy of player 1 with probability at least $1 - \eta$.  This will prove the Lemma.

To prove the claim, suppose for contradiction's sake that no such strategy for player 2 exists.  Then, by the minimax theorem, there exists a randomized strategy for player 1 which wins with probability greater than $\eta$ against all choices of $x$.  This strategy is itself a randomized algorithm making at most $\eta R_0(f)$ queries; let us call this algorithm $\mathcal{R}$.  Consider the algorithm $\mathcal{R}'$ for $f$ that on input $x$, repeatedly applies $\mathcal{R}$ to $x$ until $\mathcal{R}$ produces an output, which $\mathcal{R}'$ then outputs.  We have $\mathcal{R}'(x) = f(x)$ on every input. 
Also, the expected number of queries of $\mathcal{R}'$ on any input is strictly less than
\begin{align*}
\sum_{m \geq 1}(1 - \eta)^{m - 1}\eta \left(m\cdot \eta R_0(f) \right) \ &= \ \left(\sum_{m \geq 1}(1 - \eta)^{m - 1}m \right)\cdot \eta^2 R_0(f) \\
&= \ \frac{1}{\eta^2}\cdot \eta^2R_0(f)  \\
&= \ R_0(f), 
\end{align*} 
contradicting the definition of $R_0(f)$.
\end{proof}

\begin{theorem}\label{wczeroerrorsdpt}  For any (not necessarily Boolean) function $f$, and $\alpha \in (0, 1/2]$, any errorless heuristic for $f^{\otimes k}$ using at most $\alpha^2 R_0(f) k/4$ queries has worst-case success probability less than $\left(  7 \alpha \ln(1/\alpha)    \right)^{k}$.   
\end{theorem}

\begin{proof}  Set $\gamma := \alpha /2$.  Let $\mu_{\gamma}$ be the distribution given by Lemma~\ref{heurlem}, so that $\suc{\gamma R_0(f)}{\mu_{\gamma}}^{\zerr}(f) \leq \gamma$.  By Theorem~\ref{zeroerrorsdpt} applied to $\alpha$, with $T := \gamma R_0(f)$ and $\eps := 1 - \gamma$,
\[    \suc{\alpha (1 - \gamma) \gamma R_0(f) k}{\mu_{\delta}^{\otimes k}}^{\zerr}(f^{\otimes k}) \ \leq \ \Pr_{Y \sim B_{k, \gamma}}\left[Y >    (1 - \alpha (1 - \gamma))k\right].     \]
We have $\alpha^2 R_0(f) k/4 \leq \alpha (1 - \gamma) \gamma R_0(f) k$ (using $\gamma \leq 1/2$), so that
\begin{align*}
\suc{\alpha^2 R_0(f) k/4}{\mu_{\gamma}^{\otimes k}}^{\zerr}(f^{\otimes k}) \ &\leq \ \Pr_{Y \sim B_{k, \gamma}}\left[Y >    (1 - \alpha(1-\gamma))k\right]\\
&< \ \left[   1 - (1 - \gamma) +  6 \alpha \ln(1/\alpha) (1 - \gamma) )   \right]^{k}\\
\end{align*}
(applying Lemma~\ref{smallchern}, with $\beta := \alpha \leq 1/2$ and $\delta := (1 - \gamma)$)
\begin{align*}
& \quad{} \quad{}  \ \ \ < \ \left(   \alpha/2 +  6 \alpha \ln(1/\alpha)    \right)^{k}\\
&\quad{} \quad{}  \ \ \ < \  \left(  7 \alpha \ln(1/\alpha)    \right)^{k}.
\end{align*}\end{proof}

\section{A direct product theorem for decision tree size}\label{sizesec}

We measure the \emph{size} of a decision tree $\mathcal{D}$, denoted $\size(\mathcal{D})$, as the number of leaf (output) vertices.  Note that this is at least $1/2$ the total number of vertices.
Define $\suc{T}{\mu}^{\size}(f)$ as the maximum success probability of any size-$T$ decision tree attempting to compute $f$ on an input drawn from distribution $\mu$.  We have the following DPT for size-bounded query algorithms:

\begin{theorem}\label{sizesdpt} Let $f$ be a Boolean function.  Suppose $\suc{T}{\mu}^{\size}(f) \leq 1 - \eps$.  Then for $0 < \alpha \leq 1$, $\suc{T^{\alpha \eps k}}{\mu^{\otimes k}}^{\size}(f^{\otimes k}) \leq  2^{\alpha \eps k}(1 - \eps)^k$.
\end{theorem}

Note how the size bound grows exponentially, rather than linearly, in $k$ in the above statement.  It is natural to expect such a statement, since the $k$-fold application of a size-$T$ decision tree is described by a size-$T^k$ decision tree.  Also note that, by convexity, Theorem~\ref{sizesdpt} also bounds the success probability of any ``randomized size-$T^{\alpha \eps k}$ algorithm'' $\mathcal{R}$, i.e., of any probability distribution over size-$T^{\alpha \eps k}$ decision trees.

\begin{proof}  The proof follows that of Theorem~\ref{mainsdpt}, except that we need a new way to quantify the resources used by each of the $k$ inputs.  First we develop some definitions pertaining to a single input to $f$. Given $u \in \{0, 1, *\}^n$ and a real number $Z \in [1, T]$, let 
\[W_{\size}(u, Z) \ := \ \suc{Z}{\mu^{(u)}}^{\size}(f).\] 

\begin{lemma} \label{sizeWlem} 1. $W_{\size}(*^n, T) \leq 1 - \eps$.  

\vspace{.5 em}

2. Take any real numbers $S^{(0)}, S^{(1)} \geq 1$ and let $S := S^{(0)} + S^{(1)}$. Then for any $u \in \{0, 1, *\}^n$ and any $i \in [n]$,
\[ \bE_{\mathbf{y} \sim \mu^{(u)}}[ W_{\size}(u[x_i \leftarrow \mathbf{y}_i], S^{(\mathbf{y}_i)})  ]  \ \leq \ W_{\size}(u, S).\]  
\end{lemma}

The proof is very similar to that of Lemma~\ref{Wlem}, and is omitted.

Now let $\mathcal{D}$ be any deterministic algorithm of size at most $T^{\alpha \eps k}$ attempting to compute $f^{\otimes k}$ on input strings $\mathbf{x} = (\mathbf{x}^1, \ldots, \mathbf{x}^k) \sim\mu^{\otimes k}$.  Let $M := \lfloor T^{\alpha \eps k}\rfloor$; $\mathcal{D}$ always makes at most $M$ queries.

As in previous proofs, for $j \in [k]$ and $0 \leq t \leq M$, let $u_t^{j} \in \{0, 1, *\}^n$ describe the outcomes of all queries made to $\mathbf{x}^j$ after $\mathcal{D}$ has taken $t$ steps (here a ``step'' consists of a query, unless $\mathcal{D}$ has halted, in which case a step has no effect).  

Let $S_t$ be defined as the size (number of leaf vertices) of the subtree of $\mathcal{D}$ reached after $t$ steps have been taken.  Thus we have $S_0 \leq T^{\alpha \eps k}$, and $S_t = 1$ iff $\mathcal{D}$ has halted after at most $t$ queries.  For each $j \in [k]$, we define a sequence $Z_{j, 0}, \ldots, Z_{j, M}$, as follows.  Let $Z_{j, 0} := T$.  For $0 \leq t < M$, if $\mathcal{D}$ has halted after $t$ steps, let $Z_{j, t+1}:= Z_{j, t}$.  Otherwise, if the $(t + 1)$-st query made by $\mathcal{D}$ is \emph{not} to $\mathbf{x}^j$, we again let $Z_{j, t+1} := Z_{j, t}$.  If the $(t + 1)$-st query is to $\mathbf{x}^j$, let 
\[Z_{j, t+1} \ := \ \frac{S_{t+1}}{S_t}\cdot Z_{j, t}.\]

Let $X_{j, t} :=W_{\size}(u_j^t, Z_{j, t})$ if $Z_{j, t} \geq 1$; otherwise let $X_{j, t} := 1/2$.  
Let $P_t := \prod_{j \in [k]}X_{j, t}$. 
Arguing as in Theorem~\ref{mainsdpt}, for each $0 \leq t < M$, $\bE[P_{t+1}] \leq \bE[P_{t}]$.  
It follows that $\bE[P_M] \leq \bE[P_0] = W_{\size}(*^n, T)^k \leq (1 - \eps)^k$.  

Condition on any complete execution of $\mathcal{D}$, as described by $u^1_M, \ldots, u^k_M$.  Notice that if $Z_{j, M} \geq 1$, then (by the definitions) $X_{j, M}$ is an upper bound on the conditional success probability of guessing $f(\mathbf{x}^j)$ correctly.  Also, $X_{j, t} \geq 1/2$ for all $j, t$, and all inputs are independent after our conditioning.  Thus the conditional success probability of computing $f^{\otimes k}(\mathbf{x})$ is at most $2^{|B|}\cdot P_M$, where we define the (random) set $B := \{j \in [k]: Z_{j, M} < 1\}$.

Observe that $S_M = 1$, since the algorithm halts after at most $M$ steps.  Then,
\begin{align*}
1 \ = \ S_M \ &=  \ \left( \frac{S_1}{S_0}\right) \cdot \ldots \cdot \left(\frac{S_{M}}{S_{M-1}}\right) \cdot S_0 \\
&\leq \ \left( \frac{\prod_{j \in [k]} Z_{j, M}}{T^k}   \right) \cdot  T^{\alpha\eps k} \\
&\leq \ T^{-|B|} \cdot T^{\alpha\eps k}.
\end{align*}
Thus, $|B| \leq \alpha \eps k$ always.  So the \emph{overall} success probability is at most  $\bE[2^{|B|}P_M] \leq 2^{\alpha \eps k}\bE[P_M] \leq \left(2^{\alpha \eps}(1 - \eps)\right)^k$.
\end{proof}

One can also prove variants of our XOR lemma and other results in which we impose bounds on decision tree size rather than number of queries.  We omit the details.

\section{DPTs for dynamic interaction}\label{generalsec}

So far, all of the computational tasks we have studied have involved algorithms querying a collection of fixed input strings.  However, in many situations in computer science it is natural to consider more general problems of \emph{interaction} with dynamic, stateful entities.  An algorithm can still ``query'' these entities, but these actions may influence the outcomes of future queries.  In this section we describe how our proof methods can yield DPTs for these more general problems.  The methods involved are essentially the same as in previous sections, and the theorem we give is just one example of the kind of DPT we can prove for dynamic interaction, so we will only sketch the proofs here, indicating the novel elements.

We will propose a self-contained model of dynamic interaction.  We make no claims of conceptual novelty for this model, however.  Dynamic interaction has been an important concept for cryptography; in this context, \cite{Mau} proposed a model of \emph{random systems} that generalizes our model.  All of our work in this section could in principle be carried out in the random systems framework; we choose to use a different model that is somewhat simpler and adequate to our needs, and that preserves a clear resemblance to our work in previous sections. 

Much of the work in the random systems framework studies various kinds of \emph{composition} of random systems; this work aims to understand how cryptographic primitives can be combined into more complex protocols.  In this vein, \cite{MPR07} proved a result (see their Lemma 6) that can be informally described as follows: if an agent is playing games with two or more independent, non-communicating entities, then the maximum joint-success probability is achieved by following independent strategies on the different games.  This result establishes an ``ideal'' direct product property for interaction tasks with $k$ independent entities, in which the number of queries to each entity is fixed in advance.  By contrast, our focus will be on proving DPTs for query algorithms that can adaptively reallocate queries between the $k$ entities.

Now we formally define the type of entity with which our query algorithms interact.  Define an \emph{interactive automaton (IA)} as a 5-tuple 
\[\mathcal{M} \ = \ (\seeds, \states, \queries, R, \Delta),  \text{ where:}\]
\begin{itemize}
\item $\seeds, \states, \queries$ are each finite sets, and $\states$ contains a distinguished \emph{start state} $s_0$;
\item $R : \seeds \times \states \times \queries \rightarrow \{0, 1\}$ is a \emph{response mapping};
\item $\Delta : \seeds \times \states \times \queries \rightarrow \states$ is a \emph{transition mapping}.
\end{itemize}
These automata are deterministic, but we can incorporate randomness by providing random bits as part of $\seeds$.

We consider the scenario in which $\mathcal{M}$ is initialized to some seed $z \in \seeds$ according to a distribution $\mu$, along with the start-state $s_0$.  The automaton retains the value $z$ throughout an interaction with a query algorithm $\mathcal{R}$ (which does not know the value $z$), but changes its state-value.  If $\mathcal{R}$ selects the query $q \in Q$ while $\mathcal{M}$ has internal state $(z, s) \in \seeds \times \states$, then $\mathcal{M}$ returns the value $R(z, s, q)$ to $\mathcal{R}$ and transitions to the state $(z, \Delta(z, s, q))$.\footnote{We can now sketch the modeling differences between our work and \cite{Mau}.   Maurer's ``random systems'' are modeled as inherently randomized; they may or may not be finite-state machines; and they are specified ``behaviorally'' by their conditional distributions over query responses, conditioned on all possible conversation transcripts.}

There are several kinds of tasks one can associate with an IA.  One such task for the query algorithm $\mathcal{R}$ is to try to output a value $b \in B$ that satisfies some predicate $P(z, b)$, where $z$ is the seed to $\mathcal{M}$ and $P \subseteq \seeds \times B$ is a total relation over $\seeds$ and a finite set $B$.  This, of course, is a generalization of the relation problems we studied in Section~\ref{gentsec}, and it is natural to study the $k$-fold setting, in which $\mathcal{R}$ interacts with $k$ IAs, querying one of them at each step.  We assume that each IA only updates its state or sends a response to $\mathcal{R}$ when it is queried.  In particular, the IAs do not communicate with each other.

We can transform the IA interaction scenario into an equivalent one which highlights the similarity with the standard query model, and makes it easy to apply our previous work to obtain a DPT.  For simplicity assume $|\seeds| = 2^m$.  Given an IA $\mathcal{M}$ and an integer $N>0$, for each $z \in \seeds$ we define a string $\xi(z) \in \{0, 1\}^{m + (|\queries| + 1)^N  }$.  There are two types of entries in this string.  First there are $m$ ``ID'' entries, which simply contain a binary encoding of $z$.  Next there are $(|\queries| + 1)^N$ ``response'' entries,
with each such entry indexed by an $N$-tuple $\overline{q} = (q_1, \ldots, q_N) \in \left(\queries \cup \  \{*\}\right)^{N}$.  We are only interested in response-entries of form $\overline{q} = (q_1, \ldots, q_r, *, *, \ldots,*)$, where $q_1, \ldots, q_r \in \queries$.
For such an entry we define $\xi(z)_{\overline{q} } \in \{0, 1\}$ as the result of the following experiment: initialize $\mathcal{M}$ to state $(z, s_0)$, and perform the interaction in which a query algorithm asks queries $q_1, \ldots, q_r$ in that order.  Let $\xi(z)_{\overline{q} }$ be the final, $r$-th response made by $\mathcal{M}$.

Define a total relation $P_{\xi} \subseteq  \{0, 1\}^{m + (|\queries| + 1)^N  } \times B$ by
\[    P_{\xi} \ :=  \  \{(\xi(z), b): z \in \seeds \wedge \  P(z, b)      \}.   \]   
Also, given a distribution $\mu$ over $\seeds$, define $\mu_{\xi} \sim \xi(z)$, where $z \sim \mu$.
In this way we map an IA interaction task onto a relation problem of the type studied in Section~\ref{gentsec}, with a corresponding map from initialization distributions to input distributions.

A standard query algorithm $\mathcal{R}$ (as studied in all previous sections) can faithfully simulate an interaction with $\mathcal{M}$ initialized to an unknown $z \in \seeds$, if given query access to $\xi(z)$.  This works in the natural way: if its simulated queries up to the $r$-th step are $q_1, \ldots, q_r$, then for its $r$-th query to $\xi(z)$, $\mathcal{R}$ looks at the entry $(q_1, \ldots, q_r, *, *, \ldots,*)$ to learn $\mathcal{M}$'s $r$-th response.  Call an algorithm ``interaction-faithful'' if its sequence of queries to any input string always obeys this format.

Of course, not all algorithms are interaction-faithful.  For example, an unfaithful algorithm could simply look at the ID-entries to learn $z$.  Thus the relation problem $(P_{\xi}, \mu_{\xi})$ can be much easier than the IA interaction problem defined by $(\mathcal{M}, P, \mu)$.
However, if we restrict attention to the class of interaction-faithful algorithms $\mathcal{R}$, then it is not hard to see that there is an \emph{exact} correspondence between the ``difficulty'' of the two problems, at least for interactions lasting at most $N$ steps.  That is, for $T \leq N$, there is a $T$-query IA-interaction algorithm for $(\mathcal{M}, P, \mu)$ with success probability $p$, if and only if there is a $T$-query interaction-faithful standard algorithm for $(P_{\xi}, \mu_{\xi})$ with success probability $p$.  

The good news is that we can prove a DPT for interaction-faithful query algorithms in almost exactly the same way as for unrestricted query algorithms.  In fact, it's most natural to prove a DPT for a more general notion of faithfulness, which we define next.  Say we are given $n > 0$ and a map $\tau: \{0, 1, *\}^n \rightarrow \{0, 1\}^n$, called a \emph{query-restriction map}.  Say that a (standard) query algorithm $\mathcal{R}$ on $n$ input bits is \emph{$\tau$-faithful} if for every execution of $\mathcal{R}$ on any input, whenever the input bits seen by $\mathcal{R}$ seen so far are given by $u \in \{0, 1, *\}^n$, then $\mathcal{R}$ either halts, or chooses a next input bit $x_i$ to query whose index satisfies $\tau(u)_i = 1$.  In other words, a restriction map $\tau$ restricts the possible next queries which can be made by a $\tau$-faithful algorithm, in a way that depends only on the description $u$ of the bits seen so far.
Note that interaction-faithfulness as defined earlier is indeed equivalent to $\tau$-faithfulness for an appropriately-defined $\tau = \tau_{\text{int}}$.

For $k > 1$, define the \emph{$k$-fold product} of restriction map $\tau$, denoted $\tau^{\otimes k}: \{0, 1\}^{kn} \rightarrow \{0, 1\}^{kn}$, by $\tau^{\otimes k}(u^1, \ldots, u^k ) := (\tau(u^1), \ldots, \tau(u^k))$.  The map $\tau^{\otimes k}$ can be interpreted as a restriction map for algorithms making queries to a collection $x^1, \ldots, x^k$ of $n$-bit strings.  Note that $\mathcal{R}$ is $\tau^{\otimes k}$-faithful exactly if for each $j \in [k]$, $\mathcal{R}$'s queries to the $j$-th input (considered alone) are always $\tau$-faithful.
Thus, the $k$-fold IA interaction problem defined by $(\mathcal{M}, P, \mu)$ has ``difficulty'' equivalent to the $k$-fold relation problem defined by $(P_{\xi}, \mu_{\xi})$ for $\tau^{\otimes k}_{\text{int}}$-faithful algorithms, provided $N$ is chosen large enough in the definition of $\xi(\cdot)$ (relative to the query bounds we are interested in).

In light of these observations, a DPT for IA interaction algorithms follows by straightforward translation from the following DPT (generalizing Theorem~\ref{gen_tsdpt}) for standard query algorithms obeying a restriction map:  

\begin{theorem}\label{rm_gen_tsdpt}  Let $P \subseteq \{0, 1\}^n \times B$ be a total relation such that any $T$-query, $\tau$-faithful algorithm solves $P$ with probability at most $1 - \eps$ under input distribution $\mu$.

For any algorithm $\mathcal{R}$ making queries to inputs $\mathbf{x} = (\mathbf{x}^1, \ldots , \mathbf{x}^k) \sim \mu^{\otimes k}$  and producing output in $B^k$, define the random set $S[\mathbf{x}]$ as in Theorem~\ref{gen_tsdpt}.

Suppose $\mathcal{R}$ is $\tau^{\otimes k}$-faithful and $\alpha \eps Tk$-query-bounded for some $\alpha \in (0, 1]$, and $\mathcal{A}$ is any monotone subset of $\mathcal{P}([k])$.  Then conclusions 1 and 2 in Theorem~\ref{gen_tsdpt} also hold for $\mathcal{R}$.
\end{theorem}

\begin{proof} (Sketch)  The proof follows that of Theorem~\ref{gen_tsdpt}; we only describe the differences.  For $u \in \{0, 1, *\}^n$, and for a deterministic algorithm $\mathcal{D}$ on $n$ input bits, let
 \[W_{P}(u, \mathcal{D})  \ :=  \ \Pr_{\mathbf{y} \sim \mu^{(u)}}[(\mathbf{y}, \mathcal{D}(\mathbf{y})) \in P].\]  Let us say that $\mathcal{D}$ is \emph{$u$-inducing} if, on any input $x \in \{0, 1\}^n$ which extends\footnote{(as defined in Section~\ref{searchsubsec})} $u$, the outcome of $\mathcal{D}$'s first $|u|$ queries to $x$ are described by $u$.

If $|u| \leq T$, define $W_{P, \tau}(u) := \max_{\mathcal{D}} W_P(u, \mathcal{D})$, where the max ranges over all deterministic, $u$-inducing, $\tau$-faithful algorithms $\mathcal{D}$ making at most $T$ queries.  We have:

\begin{lemma} \label{rm_gentWlem} 
1. $W_{P, \tau}(*^n) \leq 1 - \eps$.  
 
\vspace{.5 em}

2. For any $u \in \{0, 1, *\}^n$ with $|u| < T$, and any $i \in [n]$ satisfying $\tau(u)_i = 1$, we have
\[\bE_{\mathbf{y} \sim \mu^{(u)}}[ W_{P, \tau}(u[x_i \leftarrow \mathbf{y}_i])  ] \ \leq \ W_{P, \tau}(u).\] 
\end{lemma}

The proof of Lemma~\ref{rm_gentWlem} follows that of Lemma~\ref{Wlem}.  The rest of the proof of Theorem~\ref{rm_gen_tsdpt} follows that of Theorem~\ref{gen_tsdpt}, with $W_{P, \tau}(u)$ taking the place of $W_{P}(u)$.
\end{proof}

One can also prove a DPT for search problems for $\tau$-faithful query algorithms, along the lines of Theorem~\ref{search_gen_tsdpt}.   When applied to interactive automata via the translation described earlier, search problems correspond to tasks whose success conditions are defined in terms of the interaction itself (rather than the hidden seed of the IA, or any output produced by the query algorithm).

\section{Questions for future work}
\begin{enumerate}
\item Can the bounds in our threshold DPTs and XOR lemma be improved?  For example, in Theorem~\ref{xorsdpt}, can one improve the success probability bound to $\frac{1}{2} \left(1 + \left[ 1 - 2\eps + O\left(\alpha\eps \right)  \right]^{k}\right)$?

\item  It is still unknown what worst-case success probability in computing $f^{\otimes k}$ can be achieved in general, when the number of queries allowed is $\alpha R_2(f) k$ for $\alpha \gg 1$.  The corresponding question in the quantum query model was settled by \cite{BNRdW}.  As mentioned earlier, $O(R_2(f) k \log k)$ queries always suffice to compute $f^{\otimes k}$ with high success probability; work of \cite{FPRU} implies that we cannot do better than this by using a bounded-error randomized algorithm for $f$ in a black-box fashion.

\item Can ideas from our work be helpful in obtaining new results in other computational models?  For example, \cite{LR11} prove a threshold DPT for quantum query algorithms computing Boolean functions, where the query bound scales as $\Omega(Q_2(f) k)$.  Can we extend this to a \emph{generalized} threshold DPT, analogous to our Theorem~\ref{gen_tsdpt}?

\end{enumerate}

\subsection*{Acknowledgements}
I thank Ronald de Wolf for numerous helpful comments, and in particular for encouraging me to look at threshold DPTs.  I also thank the anonymous reviewers.

\bibliography{sdptrefs2}
\bibliographystyle{halpha}

\end{document}